\documentclass[twocolumn,showpacs,10pt,aps,prd,amssymb]{revtex4}

\newcommand{\Scri}{\mbox{$\cal J$}}
\newcommand{\aap}{Astron. Astrophys.}

\newcommand{\apjl}{Astrophy. J. Lett.}

\newcommand{\lrir}{Liv. Rev. Relativ.}
\newcommand{\jmp}{J. Math. Phys.}
\newcommand{\jcop}{J. Comp. Phys.}

\usepackage{psfig}
\usepackage{epsfig}

\begin{document}

\title{Axisymmetric core collapse simulations using characteristic
numerical relativity}

\author{Florian Siebel$^{1}$, Jos\'e A. Font$^{2}$, Ewald
M\"uller$^{1}$, Philippos Papadopoulos$^{3}$}
 
\affiliation{$^{(1)}$Max-Planck-Institut f\"ur Astrophysik,
                     Karl-Schwarzschild-Str. 1,
                     D-85741 Garching, Germany \\
	     $^{(2)}$Departamento de Astronom\'{\i}a y Astrof\'{\i}sica,
		     Edificio de Investigaci\'on, Universidad de Valencia,
		     Dr.~Moliner 50, 46100 Burjassot (Valencia), Spain \\
	     $^{(3)}$School of Computer Science and Mathematics,
                     University of Portsmouth, 
                     Portsmouth, PO1 2EG, UK}

\begin{abstract}
We present results from axisymmetric stellar core collapse simulations in
general relativity. Our hydrodynamics code has proved robust and accurate
enough to allow for a detailed analysis of the global dynamics of the 
collapse. Contrary to traditional approaches based on the 3+1 formulation
of the gravitational field equations, our framework uses a foliation based 
on a family of outgoing light cones, emanating from a regular center, and 
terminating at future null infinity. Such a coordinate system is well adapted 
to the study of interesting dynamical spacetimes in relativistic astrophysics
such as stellar core collapse and neutron star formation. Perhaps most 
importantly this procedure allows for the unambiguous extraction of 
gravitational waves at future null infinity without any approximation, 
along with the commonly used quadrupole formalism for the gravitational 
wave extraction. Our results concerning the gravitational wave signals 
show noticeable disagreement when those are extracted by computing the 
Bondi news at future null infinity on the one hand and by using the 
quadrupole formula on the other hand.
We have strong indication that for our setup the quadrupole formula on
the null cone does not lead to physical gravitational wave
signals. The Bondi gravitational wave signals extracted at infinity
show typical oscillation frequencies of about 0.5 kHz.
\end{abstract}

\date{\today}  


\maketitle

\section{Introduction}
\label{introduction}


Supernova core collapse marks the final stage of the stellar evolution 
of massive stars. If the core collapse and/or the supernova explosion 
are nonspherical, part of the liberated gravitational binding energy will 
be emitted in the form of gravitational waves. According to estimates 
from numerical simulations, the total energy emitted in gravitational 
waves in such events can be as high as $10^{-6} M_{\odot} c^2$
\cite{Mue82,MSM91,ZwM97,DFM022}. Nonsphericity can be caused by the effects of
rotation, convection and anisotropic neutrino emission leading either
to a large-scale deviation from spherical symmetry or to small-scale
statistical mass-energy fluctuations (for a review, see e.g.,
\cite{Mue98}). Supernovae have always been considered among the most 
important sources of gravitational waves to be eventually detected by the 
current or next generations of gravitational wave laser interferometers.
If detected, the gravitational wave signal could be used to probe the
models of core collapse supernovae and to study the formation of neutron
stars.


Earlier studies of axisymmetric supernova core collapse were performed using
Newton's law of gravity~\cite{MSM91,YaS94,ZwM97}. More recently, effects of
general relativity have been included under the simplifying assumption of a 
conformally flat spatial metric~\cite{DFM01,DFM02,DFM022}. In all existing 
works gravitational waves are not calculated instantly within the numerical 
simulation, but they are extracted {\it a posteriori} using the approximation 
of the quadrupole formalism which links gravitational waves to the change of 
the quadrupole moment of the simulated matter distribution.


In this paper we present first results of a project aimed at studying 
the dynamics of stellar core collapse by means of numerical simulations in 
full general relativity. The trademark of our approach is the use of the 
so-called {\it characteristic formulation} of general relativity (see~\cite{Win01} 
for a review), in which spacetime is foliated with a family of outgoing light 
cones emanating from a regular center. Due to a suitable compactification
of the global spacetime future null infinity is part of our finite numerical
grid where we can unambiguously extract gravitational waves. This remarkable
feature is the main motivation behind our particular choice of slicing and 
coordinates, which clearly departs from earlier investigations. We note, however,
that we have not modeled the detailed microphysics of core collapse supernovae,
which is beyond the scope of the present investigation. Instead, we only
take into account the most important features for both, the gravitational field
and the hydrodynamics, and those will be introduced in the upcoming section. 


Characteristic numerical relativity has traditionally focused on vacuum
spacetimes. In recent years the field has witnessed steady improvement,
and robust and accurate three-dimensional codes are nowadays available, 
as that described in~\cite{BGL97}, which has been applied to diverse studies 
of black hole physics (e.g.~\cite{GLM98}). In black hole spacetimes, only the 
geometry outside a horizon is covered by the foliation of light cones. This
is is not the case for neutron stars or gravitational collapse spacetimes which
must include a regular origin. Up to now, characteristic vacuum codes
with a regular center have only been studied in spherical symmetry and in
axisymmetry~\cite{GPW94}. 

The inclusion of relativistic hydrodynamics into the characteristic
approach along with the implementation of high-resolution shock-capturing 
(HRSC) schemes in the solution procedure was first considered by Papadopoulos 
and Font~\cite{PaF00,PaF992,Fon00}. First applications in spherical symmetry 
were presented, dealing with black hole accretion~\cite{PaF99} and the 
interaction of relativistic stars with scalar fields~\cite{SFP02} as simple 
models of gravitational waves. Axisymmetric studies of the Einstein-perfect fluid
system were first discussed in~\cite{SFM02}. In this reference we presented
an axisymmetric, fully relativistic code which could maintain
long-term stability of relativistic stars and which allowed us to perform mode-frequency
computations and the gravitational wave extraction of perturbed stellar
configurations. The core collapse simulations presented in the current
paper are based on this code, which is described in detail in Ref.~\cite{SFM02}.


The paper is organized as follows: In Sec.~\ref{sec:framework} we describe 
the mathematical and numerical framework we use in the simulations.  
Section~\ref{sec:models} deals with presenting the initial data of the unstable 
equilibrium stellar configurations we evolve. Section~\ref{sec:dynamicsSN} 
is devoted to discuss the numerical simulations, with emphasis on the 
collapse dynamics. In Sec.~\ref{sec:GWfromSN} we analyze the corresponding 
gravitational wave signals. A summary and a discussion of our results are 
given in Sec.~\ref{sec:discussion}. Finally, tests to calibrate the code 
in simulations of core collapse are collected in the Appendix.

\section{Framework and implementation}
\label{sec:framework}

We only briefly repeat the basic properties of our approach here. The interested
reader is referred to our previous work~\cite{SFM02} for more details concerning 
the mathematical setup and the numerical implementation. As described in 
Ref.~\cite{SFM02}, we work with the coupled system of Einstein and relativistic 
perfect fluid equations
\begin{eqnarray}
G_{ab} & = & \kappa T_{ab} \,, \\
\label{fluid}
\nabla_{a} T^{ab} & = & 0 \ ,\\
\label{continuity}
\nabla_{a} J^a & = & 0,
\end{eqnarray}
where $\nabla_a$, as usual, denotes the covariant derivative.
The energy-momentum tensor for a perfect fluid $T_{ab}$ takes the form.
\begin{equation}
T_{ab} = \rho h u_{a} u_{b} + p g_{ab}.
\end{equation}
Here $\rho$ denotes the rest mass density, $h = 1 + \epsilon + \frac{p}{\rho}$ is 
the specific enthalpy, $\epsilon$ is the specific internal energy, and $p$ is the 
pressure of the fluid. The four-vector $u^{a}$, the 4-velocity of the fluid, fulfills 
the normalization condition $g_{ab} u^{a} u^{b} = -1$.  The four-current $J^a$ is 
defined as $J^a = \rho u^a$. Using geometrized units ($c=G=1$) the coupling constant 
in the field equations is $\kappa = 8 \pi$. We further use units in which $M_{\odot}=1$. 
Moreover, an equation of state (EoS) needs to be prescribed, $p=p(\rho, \epsilon)$, as
we discuss in Sec.~\ref{subsec:EoS} below. 

Our numerical implementation of the field equations of general relativity is based on a
spherical null coordinate system $(u,r,\theta,\phi)$. Here, $u$ denotes a null coordinate 
labeling outgoing light cones, $r$ is a radial coordinate, and $\theta$ and $\phi$ are 
standard spherical coordinates. Assuming axi\-symmetry, $\phi$ is a Killing coordinate. 
In order to resolve the entire radial range from the origin of the coordinate system up to
future null infinity, we define a new radial coordinate $x \in[0,1]$. The radial coordinate 
$r$ is a function of the coordinate $x$, which can be adapted to the particular simulation. 
In this work, except where otherwise stated,  we use a grid function 
$r(x)=100\tan({\frac{\pi}{2}x})$, for which the limit $x\to1$ corresponds to $r\to\infty$. 
Moreover, in order to eliminate singular terms at the poles ($\theta=0,\pi$) we introduce the new
coordinate $y = -\cos \theta$.

\subsection{The characteristic Einstein equations} 

The geometric framework relies on the Bondi (radiative) metric~\cite{BBM62}
\begin{eqnarray}
\label{Bondi}
& & ds^{2} = -\left(\frac{V}{r}e^{2 \beta} - U^{2} r^{2} e^{2 \gamma}\right) du^{2} 
 - 2 e^{2 \beta} du dr 
\nonumber \\ 
 &-& 2 U r^{2} e^{2 \gamma} du d\theta 
 + r^{2}
 (e^{2 \gamma} d\theta^{2} + e^{-2 \gamma} \sin^{2}\theta d\phi^{2}).\\
\nonumber
\end{eqnarray}
We substitute the metric variables $(\beta,V,U,\gamma)$ by the new set
of metric variables $(\beta,S,\hat{U},\hat{\gamma})$,
\begin{eqnarray}
S & = & \frac{V - r}{r^{2}} , \\
\label{Uhat}
\hat{U} & = & \frac{U}{\sin \theta}, \\
\nonumber\\
\label{hatgamma}
\hat{\gamma} & = & \frac{\gamma}{\sin^{2} \theta},
\end{eqnarray}
in order to obtain regular expression in the Einstein equations in
particular at the polar axis.
The origin of the coordinate system $r=0$ is chosen to lie on the axis 
of our axisymmetric stellar configurations, where we describe
boundaries and falloff conditions for the metric fields. The complete
set of Einstein equations reduces to a wave equation for the
quantity $\hat{\gamma}$ (see Eq.~(\ref{hatgamma})) and a hierarchical set of 
hypersurface equations for the quantities $(\beta,\hat{U},S)$ 
to be solved along the light rays $u=\text{const}$. The
particular form of these equations is explicitly given in
Ref.~\cite{SFM02}.

\subsection{The relativistic perfect fluid equations}

The axisymmetric general relativistic fluid equations on the light cone, 
Eqs.~(\ref{fluid}) and~(\ref{continuity}), are written as a first-order 
flux-conservative, hyperbolic system for the state-vector
${\bf U}=(U^{u},U^{x},U_{y},U^4)=(T^{uu},T^{ux},T^{u}_{~y},J^u)$. 
Following our previous work~\cite{SFP02} we have not included the metric
determinant in the definition of the state vector. Explicitly, in the
coordinates $(x^0,x^1,x^2,x^3) = (u,x,y,\phi)$, we obtain
\begin{eqnarray}
\label{conservation}
\partial_{0} U^{u} + \partial_{j} F^{ju}    & = & S^{u} \ , \\
\partial_{0} U^{x} + \partial_{j} F^{jx}    & = & S^{x} \ , \\
\partial_{0} U_{y} + \partial_{j} F^{j}_{~y} \ & = & S_{y} \ , \\
\partial_{0} U^{4} + \partial_{j} F^{j4}    & = & S^{4}.
\end{eqnarray}
The flux vectors are defined as
 \begin{eqnarray}
 F^{ju} & = & T^{ju} \ , \\
 F^{jx} & = & T^{jx} \ ,\\
 F^{j}_{~y} & = & T^{j}_{~y} \ ,\\
 F^{j4} & = & J^{j} \ ,
 \end{eqnarray}
 and the corresponding source terms read
\begin{eqnarray}
S^{a} & = &  - \big(\ln (\sqrt{-g})\big)_{,b} \ T^{ab} +
g^{ab} \ S_{b} + T^{c}_{~b} (g^{ab})_{,c} , \hspace{10pt} \\
S_{a} & = & - \frac{1}{2} \rho h u_{c} u_{d} (g^{cd})_{,a} + p -
\big(\ln (\sqrt{-g})\big)_{,b} \ T^{b}_{~a} \ , \hspace{10pt} \\
S^{4} & = &  - \big(\ln (\sqrt{-g})\big)_{,b} \ J^{b} \ , \hspace{10pt} 
\end{eqnarray}
wherein a comma is used to denote a partial derivative.
  
The fluid update from time $u^n$ to $u^{n+1}$ at a given cell $i,j$ is given
by
\begin{eqnarray}
{\mathbf U}_{i,j}^{n+1}={\mathbf U}_{i,j}^{n}
&-&\frac{\Delta u}{\Delta x}
(\widehat{{\mathbf F}}_{i+1/2,j}-\widehat{{\mathbf F}}_{i-1/2,j}) 
\nonumber \\
&-&\frac{\Delta u}{\Delta y}
(\widehat{{\mathbf G}}_{i,j+1/2}-\widehat{{\mathbf G}}_{i,j-1/2}) 
\nonumber \\
&+& \Delta u {\mathbf S}_{i,j} \,,
\end{eqnarray}
\noindent
where the numerical fluxes, $\widehat{{\mathbf F}}$ and $\widehat{{\mathbf G}}$, 
are evaluated at the cell interfaces according to a flux-formula, the
one due to Harten, Lax and van Leer (HLL) in our case
\cite{HLL83}. The characteristic information of the Jacobian matrices
associated with the hydrodynamical fluxes, which is used in this flux 
formula, was presented 
elsewhere~\cite{PaF99}. We use the monotonized central difference slope
limiter by van Leer~\cite{vLe77} for the reconstruction of the hydrodynamical
quantities at the cell interfaces needed in the solution of the Riemann
problems. This scheme is second order accurate in smooth monotonous parts of
the flow and gives improved results compared to the MUSCL scheme applied in~\cite{SFM02} 
(for an independent comparison, see~\cite{FGI02}).

\subsection{Equation of state}
\label{subsec:EoS}

We use a hybrid EoS which includes the effect of stiffening at
nuclear densities and the effect of thermal heating due to the appearance of 
shocks. Such an EoS was first considered by Janka et al.~\cite{JZM93}, and 
has been used for core collapse simulations both, using Newtonian
gravity~\cite{ZwM97,RMR98} and in general relativity under the 
assumption of conformal flatness~\cite{DFM01,DFM02,DFM022}.

In our EoS the total pressure consists of a polytropic part, which takes 
into account the contribution from the degenerate electron gas, as well as 
the nuclear forces (at high densities), and a thermal part due to the heating 
of the material by a shock, $p = p_\mathrm{p} +p_\mathrm{th}$. More precisely, 
the polytropic part follows the relation
\begin{equation}
p_p = \left\{
\begin{array}{ll}
\kappa_1 \rho^{\Gamma_1},& \text{for $\rho \le \rho_\mathrm{n}$},\\
\kappa_2 \rho^{\Gamma_2},& \text{for $\rho > \rho_\mathrm{n}$}, 
\end{array}
\right.
\label{hybridp}
\end{equation}
where we assume a nuclear density $\rho_\mathrm{n}=2 \times 10^{14}
\ \text{g} \ \text{cm}^{-3}$. For a degenerate relativistic electron gas 
$\Gamma=\Gamma_\mathrm{ini}=4/3$ and $\kappa=4.8974894 \times 10^{14}
\ [\text{cgs}]$. To model the physical processes 
which lead to the onset of the collapse, we reduce the effective 
adiabatic index from $\Gamma$ to $\Gamma_1$ setting $\kappa_1=\kappa$ 
at the initial slice. Moreover, to model the stiffening of the EoS at 
nuclear densities, we assume $\Gamma_2=2.5$. The value of the polytropic
constant $\kappa_{2}$ follows from the requirement that the pressure 
is continuous at nuclear density. The thermodynamically consistent 
internal energy distribution reads
\begin{equation}
\label{hybridepsilonp}
\epsilon_\mathrm{p} = \left\{
\begin{array}{ll}
\frac{\kappa_{1}}{\Gamma_1-1} \rho^{\Gamma_{1}-1},   
&  \text{for $\rho \le \rho_\mathrm{n}$},\\
\frac{\kappa_{2}}{\Gamma_2-1} \rho^{\Gamma_{2}-1} + E,  
&  \text{for $\rho > \rho_\mathrm{n}$}.
\end{array}
\right.
\end{equation}
The requirement that $\epsilon_\mathrm{p}$ is continuous at nuclear
density leads to
\begin{equation}
E = \frac{(\Gamma_2-\Gamma_1)\kappa_{1}}{(\Gamma_2-1)(\Gamma_1-1)} \ 
\rho_\mathrm{n}^{\Gamma_{1}-1}.
\end{equation}
For the thermal contribution to the total pressure, we assume an ideal
fluid EoS
\begin{equation}
p_\mathrm{th}=(\Gamma_\mathrm{th}-1)\rho \epsilon_\mathrm{th},
\end{equation}
with an adiabatic index $\Gamma_\mathrm{th}=\frac{3}{2}$ describing a
mixture of relativistic and non-relativistic gases. The internal
thermal energy $\epsilon_\mathrm{th}$ is simply obtained from 
\begin{equation}
\label{hybridepsilonth}
\epsilon_\mathrm{th} = \epsilon - \epsilon_\mathrm{p}.
\end{equation}
We can summarize the EoS in a single equation:
\begin{eqnarray}
\label{hybrid}
p &=& \kappa \left(1-\frac{\Gamma_\mathrm{th}-1}{\Gamma-1}\right) \rho^{\Gamma} +
(\Gamma_\mathrm{th}-1) \rho \epsilon 
\nonumber \\
&-& \frac{(\Gamma_\mathrm{th}-1)(\Gamma-\Gamma_1)}
{(\Gamma_2-1)(\Gamma_1-1)} \kappa \rho_\mathrm{n}^{\Gamma-1} \rho,
\end{eqnarray}
where $\Gamma$ and $\kappa$ change discontinuously at nuclear density
$\rho_\mathrm{n}$ from $\Gamma_1$ to $\Gamma_2$ and $\kappa_1$ to $\kappa_2$. 
For the sound speed $c_s$, we obtain
\begin{equation}
h c_s^2 =
\frac{1}{\rho} (\Gamma p_\mathrm{p} + \Gamma_\mathrm{th} p_\mathrm{th}).
\end{equation}

\subsection{Recovery of the primitive variables}

After the time update of the state-vector of hydrodynamical quantities, 
the {\it primitive} variables $(\rho,u^x,u_y,\epsilon)$ have to be recomputed. The
relation between the two sets of variables is not in closed algebraic form.
Using the hybrid EoS, such recovery is performed as follows:
With the definition $S^{ab} = g^{cd} T^{a}_{~c} T^{b}_{~d}$, we obtain~\cite{PaF992}
\begin{equation}
\label{recover}
S^{uu} =  \Big( \frac{p}{\rho} -1 -\epsilon \Big) \Big( \frac{p}{\rho}
+1 +\epsilon \Big)
(J^u)^2 + p^2 g^{uu},
\end{equation}
where in our null coordinate system $g^{uu} = 0 $. 
Let $F(\rho,\epsilon) = \frac{p}{\rho}$. From Eqs.~(\ref{hybrid}),
(\ref{recover}) and the definition of the specific enthalpy $h$
we obtain the 3 equations for the 3 unknowns $F$, $\rho$ and $\epsilon$
\begin{eqnarray}
\label{NR1}
F(\rho, \epsilon) & = & \sqrt{L + (1+\epsilon)^{2}},\\
\label{NR2}
F(\rho, \epsilon) & = & (\Gamma_{th}-1) \epsilon + G(\rho),\\
\label{NR3}
\rho & = & H (1+\epsilon+F(\rho, \epsilon)).
\end{eqnarray}
In these equations we made use of the abbreviations
\begin{eqnarray}
L & = & \frac{S^{uu}}{(J^u)^2},\\
\nonumber
G(\rho) & = & \kappa \Big(1-\frac{\Gamma_{th}-1}{\Gamma-1}\Big) 
\rho^{\Gamma-1} \\ &   & - \frac{(\Gamma_{th}-1)(\Gamma-\Gamma_1)}
{(\Gamma_2-1)(\Gamma_1-1)} \kappa \rho_n^{\Gamma-1},\\
I & = & \frac{(J^u)^2}{T^{uu}}.
\end{eqnarray}
From Eqs.~(\ref{NR1})-(\ref{NR3}) we deduce a single implicit equation
for the rest mass density $\rho$
\begin{equation}
\label{NR}
f_{\text{imp}}(\rho) := \Big( \frac{\rho}{I} \Big)^2 - 2 \frac{\rho}{I}(1+\epsilon) -L = 0,
\end{equation}
where we consider the internal energy as function of $\rho$
\begin{equation}
\epsilon = \frac{1}{\Gamma_{th}}\Big(\frac{\rho}{I} - (1+G(\rho))\Big).
\end{equation}
We solve Eq.~(\ref{NR}) for $\rho$ with a Newton-Raphson method.

\section{Initial data}
\label{sec:models}

In the final stage of the evolution of massive stars, the iron
core in the stellar center has a central density of about $\rho_c = 10^{10} 
\ \text{g} \ \text{cm}^{-3}$ when it becomes dynamically unstable to
collapse. As the pressure of the degenerate relativistic
electrons is by far the most important contribution to the total
pressure, the pressure in the iron core can be approximated by a 
$\Gamma=\frac{4}{3}$ polytropic EoS. In order to obtain an initial model 
for the iron core, we solve the Tolman-Oppenheimer-Volkoff 
equation~\cite{SFM02} with the above central density, which corresponds 
to $\rho_c = 1.62 \times 10^{-8}$ in code units ($c=G=M_{\odot} = 1$).

To initiate the gravitational collapse we set the adiabatic index 
$\Gamma_{1}$ in the hybrid EoS~(\ref{hybrid}) to a value of $1.30$, 
which mimics the softening of the EoS due to capture of electrons and 
due to the endothermic photodisintegration of heavy nuclei. 
The chosen value is within the interval range analyzed in
previous studies of rotational core collapse based on Newtonian
physics~\cite{ZwM97} and on the conformal flat metric approximation of
general relativity~\cite{DFM01,DFM02,DFM022}. 

Since rotation is not included in our current implementation, the equilibrium
initial models of the iron core are spherically symmetric. Furthermore, in
the evolution of these data during the phases of collapse, bounce, and beyond, 
spherical symmetry is conserved. Therefore, since we are mainly interested 
in simulating core collapse as a source of gravitational waves, we add non-radial
perturbations on top of the spherical data. Our analysis is thus restricted to
collapse scenarios where the effects of rotation are unimportant and in which 
stellar evolution has led to asymmetries in the iron core, e.g. due to 
convection~\cite{BaA94}. The strongest gravitational wave signals are expected for
perturbations of quadrupolar form. Hence, we further restrict our analysis to
this case, varying the form and amplitude of the perturbation in the
initial data. We note that the evolution of such data, however, can
produce an arbitrary type of perturbation within the class of the
imposed symmetry.

We have classified the different models as follows: In case
$\mathfrak{A}$ the spherical model is unperturbed; in case
$\mathfrak{B}$ we prescribe a perturbation of the rest mass density
\begin{equation}
\delta \rho = A \rho_s \sin \Big(\frac{\pi r^2}{R^2} \Big) y^2,
\label{pertB}
\end{equation}
where $\rho_s$ denotes the spherical density distribution.
Finally, in case $\mathfrak{C}$ we prescribe a perturbation of the meridional 
velocity component
\begin{equation}
u_y = A \sin \Big(\frac{\pi r^2}{R^2} \Big) y.
\label{pertC}
\end{equation}
In the above two equations $A$ is a free parameter describing the amplitude of the
perturbation, and $R$ denotes the radius of the iron core ($R = 1.4 \times
10^3$ km). We note in passing that in~\cite{SFM02} we already used a 
perturbation of the form $\mathfrak{C}$ to study quadrupolar oscillations 
of relativistic stars. We have further classified models $\mathfrak{B}$ and 
$\mathfrak{C}$ according to the amplitude $A$ of the perturbation (e.g. case 
$\mathfrak{C}01$  would correspond to an amplitude $A=0.1$).

\begin{figure}[t]
\centerline{\psfig{file=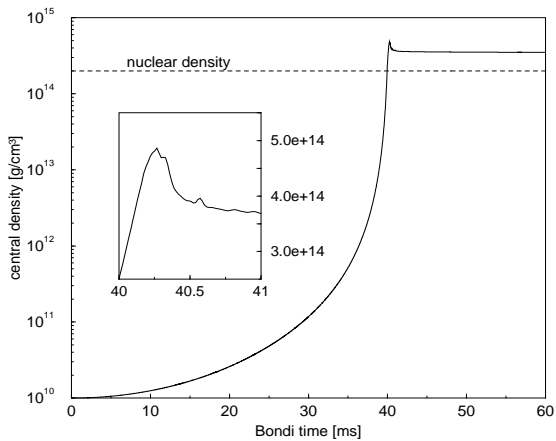,width=2.3in,height=2.9in,angle=-90}}
\caption{Evolution of the central density for the collapse model
$\mathfrak{B}01$ using a semilogarithmic scaling. During the collapse
the central density increases by 4.5 orders of magnitude. When
reaching supra-nuclear densities, the collapse is stopped as a
consequence of the stiffening in the hybrid EoS at about $40$ ms after 
the collapse was initiated. The central density finally approaches a new 
equilibrium supra-nuclear value. Shortly after bounce, oscillations appear 
in the central density (see inset).
\label{fig7.1}}
\end{figure}

\section{Core collapse dynamics}
\label{sec:dynamicsSN}

This section deals with the description of the global dynamics of our core 
collapse simulations. Relevant tests of the code which assess its suitability 
for such simulations are collected in the Appendix.

\subsection{Collapse and bounce}

When evolving the initial models described in the previous section, 
the core starts to collapse. Fig.~\ref{fig7.1} shows the
evolution of the central density for model~$\mathfrak{B}01$ as a
function of the Bondi time $u_B$. The lapse of Bondi time as seen by an
observer at infinity is defined by 
\begin{equation}
du_B = \omega e^{2H} du,
\end{equation}
where $H=\lim_{r \to \infty} \beta$.
The conformal factor $\omega$ relates the two-geometry of the Bondi metric
\begin{equation}
d \hat{s}^{2} = e^{2\gamma} d \theta^{2} + \sin^{2} \theta e^{-2\gamma} d \phi^{2}
\end{equation}
to the two-geometry of a unit sphere
\begin{equation}
d \hat{s}_{B}^{2} = d \theta_{B}^{2} + \sin^{2} \theta_{B} d\phi_{B}^{2}.
\end{equation}
as $d \hat{s}_{B}^{2} = \omega^2 d \hat{s}^{2}$.
When the central density reaches nuclear density at a Bondi time of
about 40 ms, the pressure increases strongly according to 
Eq.~(\ref{hybridp}). The central density grows further, but its increase 
is soon stopped. Afterwards, it drops below its maximum value, finally 
approaching a quasi-equilibrium supra-nuclear value when a
``proto-neutron star'' has formed in the central region~\cite{note1}.

\begin{figure}[t]
\centerline{\psfig{file=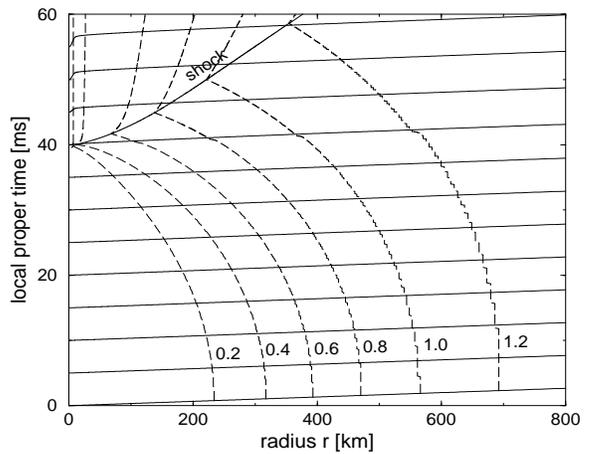,width=2.4in,height=3.0in,angle=-90}}
\caption{Spacetime diagram for the collapse model $\mathfrak{A}$.
Plotted is the lapse of proper time as a function of the radial coordinate 
$r$. The black solid lines correspond to a subset of the light curves by which
we foliate the spacetime (there is one light cone after every 5 ms, where
time is measured by an observer at the origin). The dashed curves correspond 
to different mass shells: $M=0.2 M_{\odot}, 0.4 M_{\odot}, 0.6
M_{\odot}, 0.8 M_{\odot}, 1.0 M_{\odot}, 1.2
M_{\odot}$. After about $40$ ms, a shock (thick solid line) forms in
the interior region close to the origin. The diagram was obtained from 
a global simulation with 800 radial zones, extending the grid to future null 
infinity.
}
\label{fig7.2}
\end{figure}

\begin{figure}[t]
\centerline{\psfig{file=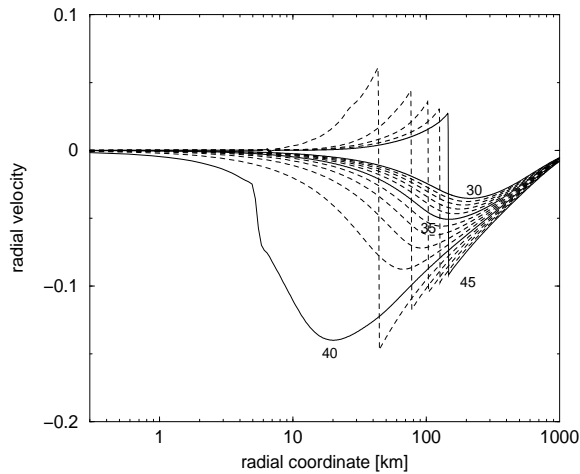,width=2.5in,height=3.0in,angle=-90}}
\caption{Snapshots of radial velocity profiles $u^r$, plotted
as function of radius $r$ for the collapse model $\mathfrak{A}$.
The snapshots are taken between $u_B=30$ ms and $u_B=45$ ms, with
a delay of $1$ ms between subsequent outputs (the solid lines
correspond to $u_B=30, 35, 40, 45$ ms). The shock formation
takes place at about 40 ms. In the outer part of the
plotted region, the infall velocity of matter increases
monotonically with time. 
}
\label{fig7.3}
\end{figure}

Fig.~\ref{fig7.2} shows a spacetime diagram for the core collapse simulation 
of model $\mathfrak{A}$ (the main aspects are similar for all our models).
The diagram shows different mass shells and the location of the shock
front (thick solid line). In order to localize the shock front, we search 
for coordinate locations where the $x$-component of the 4-velocity
$u^x$ fulfills $u^{x}_{i} - u^{x}_{i+1} \ge s$, with $s$ being 
a threshold value for a velocity jump to be adapted (typical values for our 
simulations are $s=10^{-5} ... 10^{-4}$). In addition, to compute the
mass inside a fixed radius, we make use of the relation
\begin{equation}
M = 4 \pi \int_{0}^{\infty} r^{2} e^{-2 \beta} T_{ru} dr,
\end{equation}
valid for the spherical collapse model $\mathfrak{A}$. Fig.~\ref{fig7.2} 
shows that at the beginning of the collapse phase, the spacetime metric is 
close to the Minkowski metric, which is reflected in the diagram by the light
cones being almost parallel straight lines. The effects of curvature can be 
most strongly seen close to the origin ($r=0$) after about $40$ ms, when the
proto-neutron star has formed. We observe a redshift factor $e^{2H}$ relating 
the lapse of local proper time at the origin to the lapse of proper time at
infinity of $\sim 1.12$. 

\begin{figure}[t]
\centerline{\psfig{file=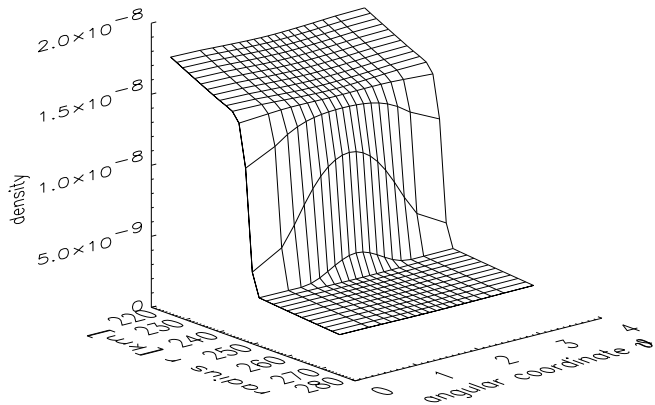,width=4.0in,height=2.7in,,angle=0}}
\caption{Surface plot of the rest mass density distribution $\rho$ around the 
shock front for the collapse model $\mathfrak{C}01$. $50$ ms after the collapse
was initiated, the shock has reached a radius of about $250$ km. We plot every 
radial zone using a radial grid $r = 100 \tan(\frac{\pi}{2}x)$ with 450 radial 
zones. The shock front is resolved with only three radial zones. The aspherical 
nature of the data is most prominent at the shock front.
}
\label{fig7.4}
\end{figure}

\begin{figure}[h]
\centerline{\psfig{file=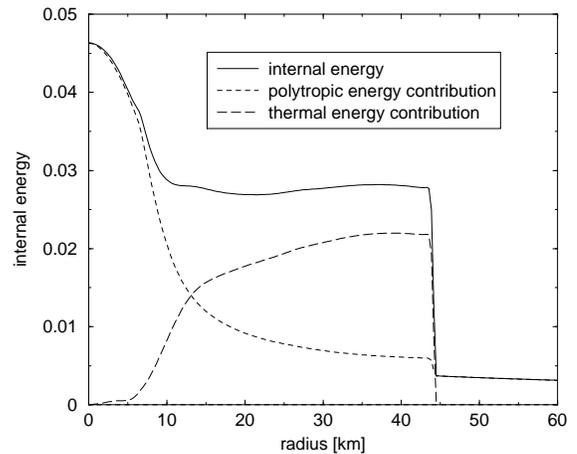,width=2.4in,angle=-90}}
\caption{Radial distribution of the internal energy $\epsilon$ (solid
line) shortly after bounce ($u_B = 41$~ms) for the collapse model 
$\mathfrak{A}$. The different contributions from the polytropic part
$\epsilon_p$ (dashed line) and the thermal part $\epsilon_{th}$ 
(long-dashed line) to the total internal energy are also shown.
In front of the shock which is located at a radius of $\sim 45$ km, 
the thermal energy vanishes.
}
\label{fig7.5}
\end{figure}

Correspondingly, Fig.~\ref{fig7.3} shows different snapshots of the radial 
velocity $u^r$ at evolution times close to bounce. In the inner region
(the so-called {\it homologous} inner core), the 
infall velocity measured as function of radius is proportional to the
radius. The homologous inner core shrinks with time. The outer limit of the homologous
region, i.e. the sonic point, where the local sound speed has the same 
magnitude as the infall velocity, finally reaches a radius of less than $10$ 
km after about $40$ ms. At that time, the shock front forms, which moves 
outwards with a speed of $\sim 0.1$c initially. During its propagation it 
is gradually slowed down by the interaction with the infalling material in 
the outer region. It is worth to stress the ability of the code to resolve 
the steep shock front within only a few grid zones (typically three). This can 
be further seen in Fig.~\ref{fig7.4}, where we plot the rest mass density 
$\rho$ at the shock front for a simulation of the collapse model 
$\mathfrak{C}01$.

Matter falling through the outward propagating shock is heated substantially.
This can be seen in
Fig.~\ref{fig7.5}, where we plot the internal energy
distribution $\epsilon$ in the central region shortly after
bounce. The figure further shows the contribution to the internal
energy from the polytropic part, Eq.~(\ref{hybridepsilonp}),
and the thermal part, Eq.~(\ref{hybridepsilonth}). In the very
central region, the polytropic contribution constitutes the dominant 
part. In contrast, the thermal energy dominates the total internal 
energy in the post-shock region for radii larger than a certain value
(the shock forms off center), $\sim 13$ km in the specific situation
shown in Fig.~\ref{fig7.5}. We have verified that the global energy balance 
(see Ref.~\cite{SFM02} for more details) is well preserved in our
simulations (maximum errors are of the order of $0.5-1 \%$).

\begin{figure}[t]
\centering
\vspace{-0.3cm}
\includegraphics[angle=90, width= 1.0 \linewidth]{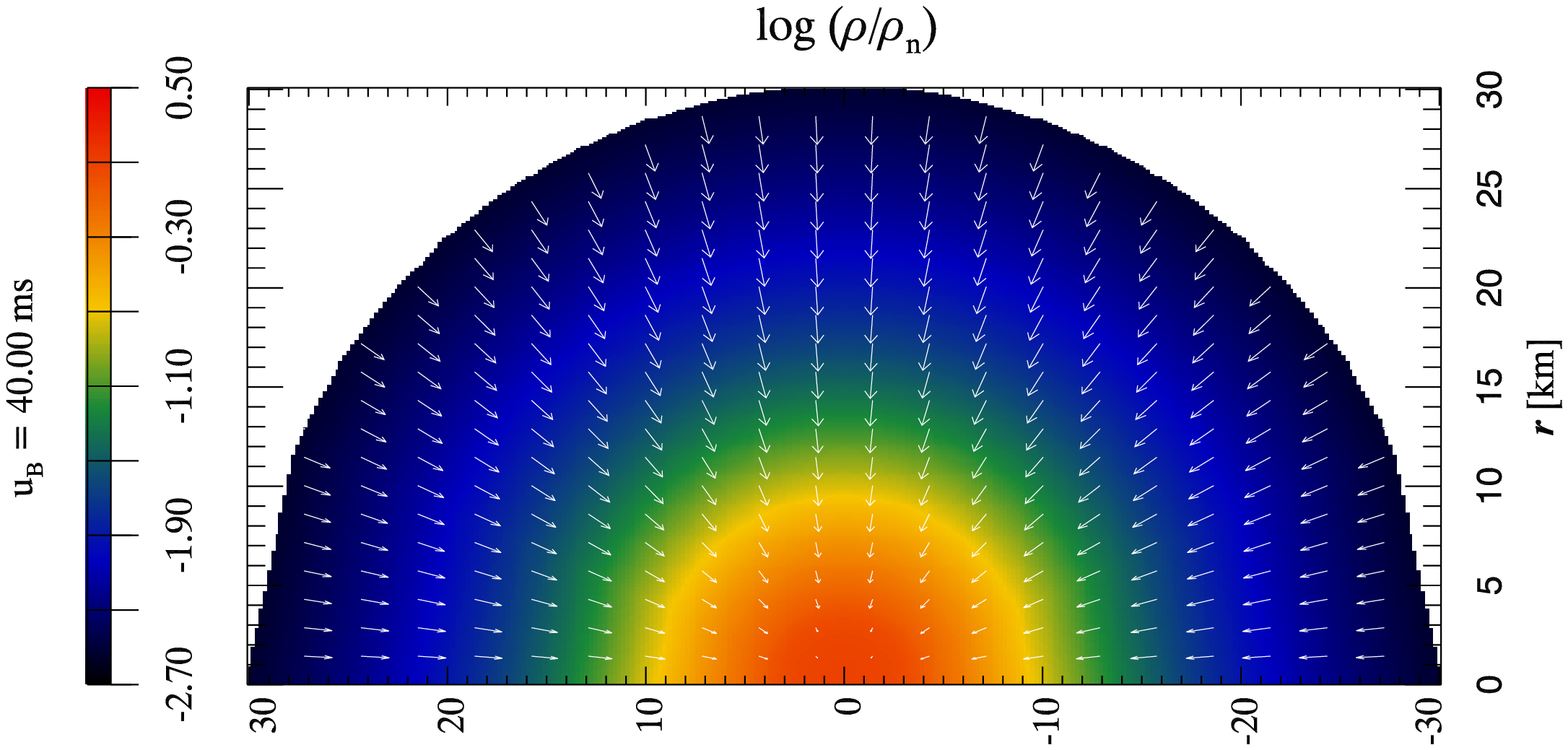}
\vspace{-0.75cm}\\
\includegraphics[angle=90, width= 1.0 \linewidth]{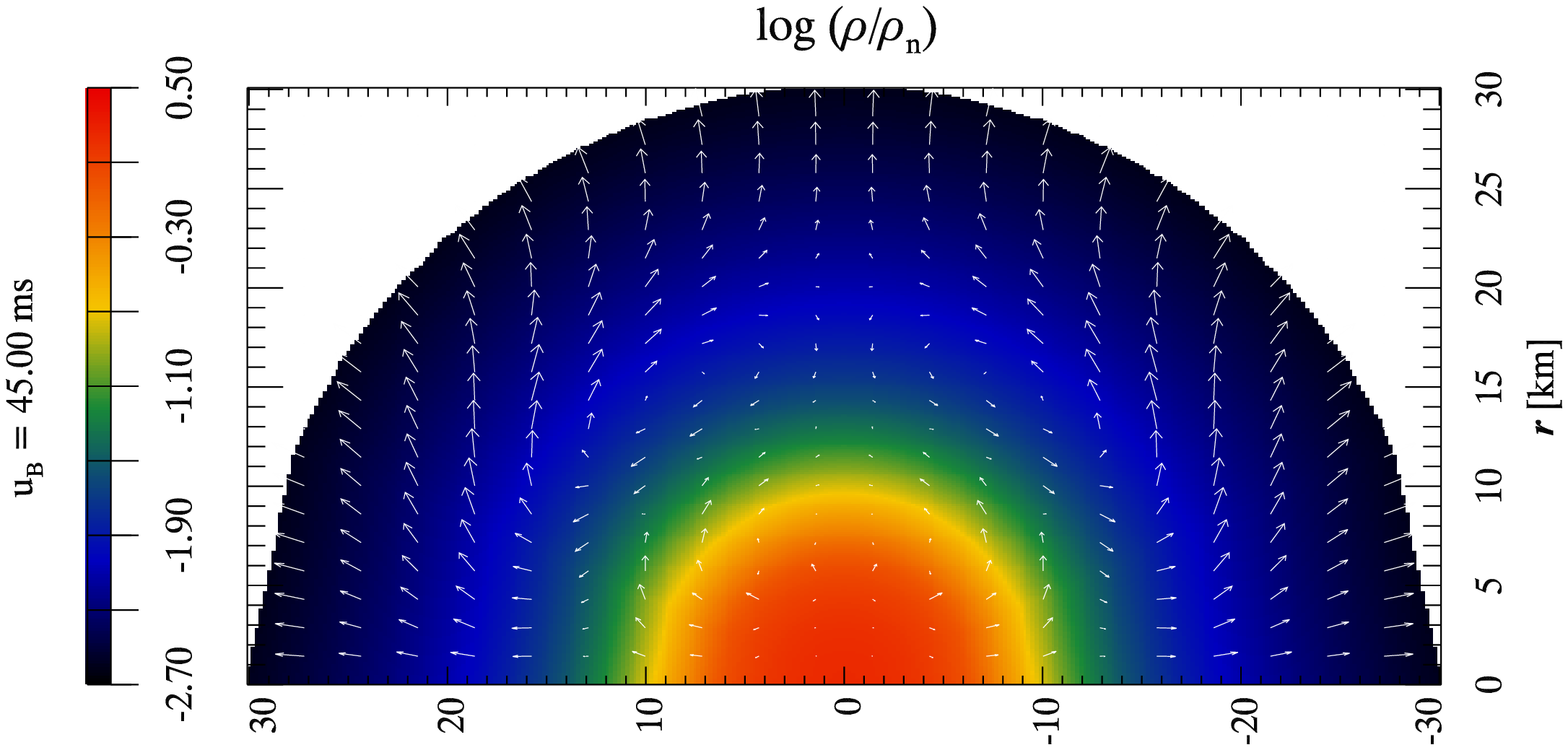}
\vspace{-0.75cm}\\
\includegraphics[angle=90, width= 1.0 \linewidth]{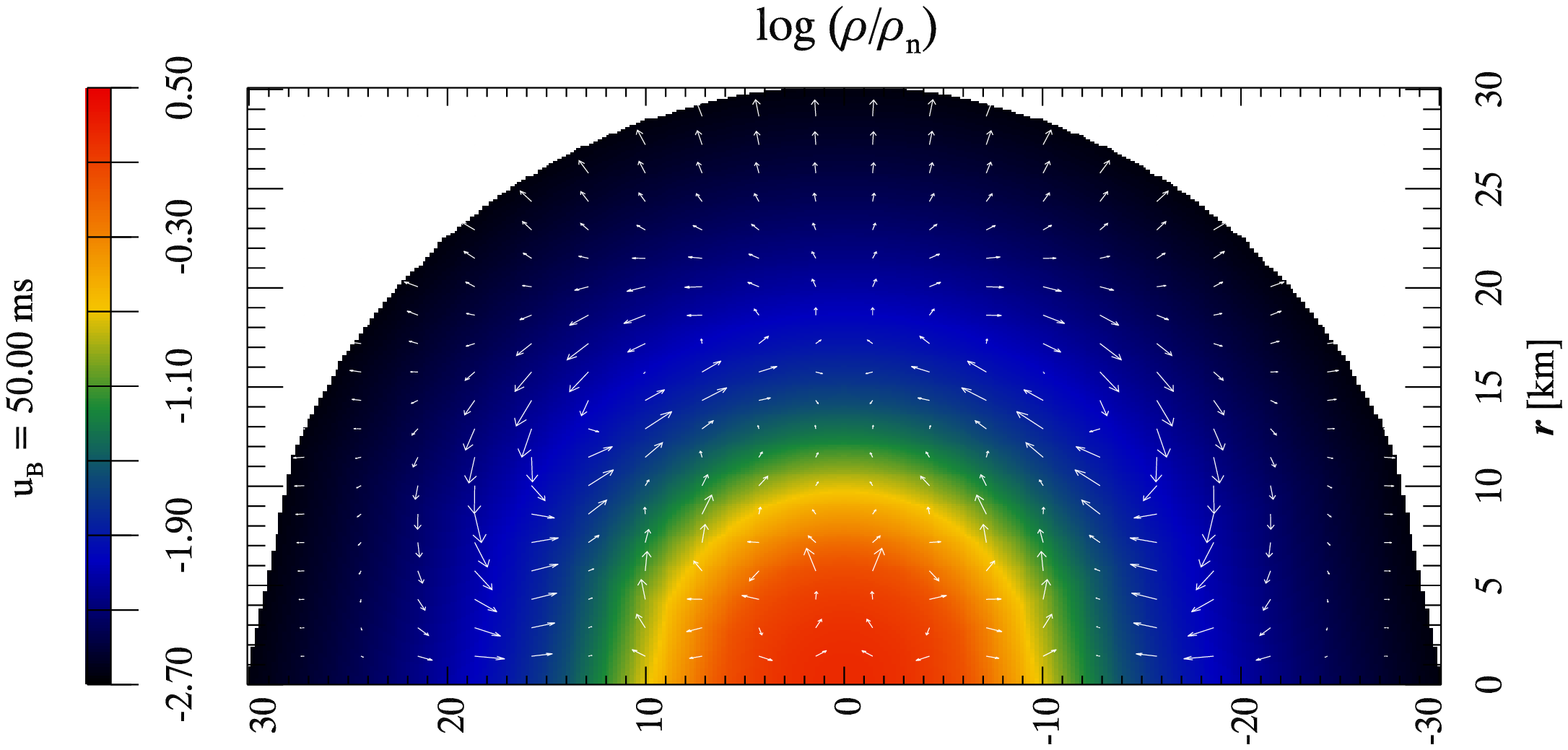}
\vspace{-0.75cm}
\caption{Contour plot of the rest mass density distribution for model
$\mathfrak{B}01$ at a Bondi time $u_B = 40$ ms (upper panel), $u_B =
45$ ms (middle panel) and $u_B = 50$ ms (lower panel), obtained from a global
evolution extending the grid to future null infinity. We only show a fraction
of the core up to a radius of $30$ km. Overlayed are velocity
vectors. At bounce (upper panel), the
matter distribution is, to a great extent, spherically symmetric . In the later
phases (middle and lower panels), the fluid dynamics is characterized
by aspherical flows related to the oscillations of the newborn neutron
star. The matter flow shows reflection symmetry with respect to the
equator, which is inherent to the initial data and well preserved
during the evolution.
\label{fig7.7}}
\end{figure}

Fig.~\ref{fig7.7} shows two-dimensional contour plots illustrating 
the dynamics during collapse and bounce for model $\mathfrak{B}01$. For 
this particular simulation we used a resolution $(N_x,N_y) = (600,12).$ 
The figure displays isocontours of the rest mass density covering only the 
inner part of the iron core up to a radius of $30$ km at $40$ ms 
(i.e. at bounce; top panel), at 45 ms (when the shock has reached a radius 
of $\sim 140$~km; middle panel) and at 50 ms (when the shock wave is 
located at $r\sim 250$~km; bottom panel). The velocity vectors overlayed onto 
the contour plots are normalized to the maximum velocity in the displayed 
region. During the collapse phase until bounce at nuclear densities (upper panel), 
the initial aspherical contributions do not play a major role - the radial infall 
velocities dominate the dynamics. After bounce (middle and lower panel) 
the newly formed neutron star in the central region shows nonspherical 
oscillations, with fluid velocities up to about $2 \times 10^{-3}c$.
Qualitatively, the dynamics for the collapse model $\mathfrak{C}01$ is
very similar to what is shown in Fig.~\ref{fig7.7} for
model $\mathfrak{B}01$. However, the particular form of the
non-spherical pulsations created after bounce differs.

\subsection{Fluid oscillations in the outer core}

\begin{figure}[t]
\centerline{\psfig{file=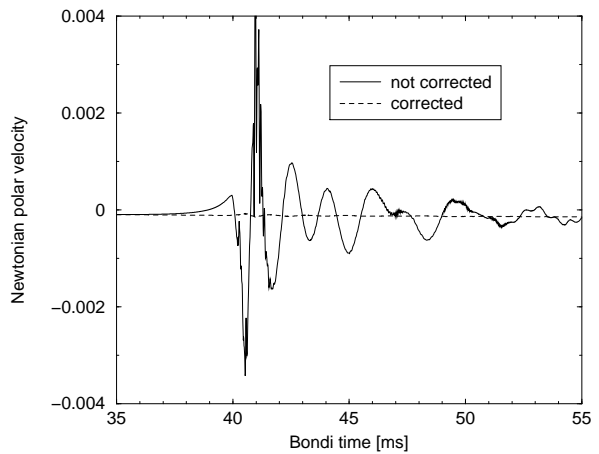,width=0.7 \linewidth,angle=-90}}
\caption{Meridional velocity component as a function of Bondi
time at the fixed location $r = 833$ km and $y=0.5$ for model
$\mathfrak{B}01$. The radial location was chosen well ahead of the
shock. The solid line corresponds to the meridional velocity as extracted 
in our coordinate system, $v_2 = r u^{\theta}$, in units of the speed of 
light $c$. The dashed line corresponds to the meridional velocity evaluated 
in inertial Bondi coordinates defined at future null infinity. See text 
for more details.}
\label{fig7.8}
\end{figure}

When analyzing the dynamical behavior of the fluid after bounce, we find 
that the meridional velocity oscillates strongly in the entire
pre-shock region. This can be seen from the solid curve of
Fig.~\ref{fig7.8}, where we plot the meridional velocity component
$v_2 = r u^{\theta}$ for model $\mathfrak{B}01$ as a function of the
Bondi time, and at coordinate location $r = 833$ km and $y=0.5$.
These oscillations are created directly after the formation of the proto-neutron 
star in the central region of the numerical domain. The only possibility 
to propagate information instantaneously (i.e.~on a slice with constant 
retarded time $u$) from the central region to the outer layers of the iron core
is through the metric, since sound waves would need several 10 ms to cover
the distance. There are two possible
explanations for these oscillations. Either they are created when
gravitational wave energy is absorbed well ahead of
the shock, or they are created by our choice of coordinates, i.e.
they are gauge effects. In the latter case, the oscillations would
not be caused by a real flow, but as a consequence of the
underlying coordinate system in which we describe the flow.

To clarify the origin of the oscillations we estimate in the following
the kinetic energy of the oscillations, assuming that they are a physical 
effect. The average amplitude of the oscillation
is of the order of $\hat{v}_2 = 2 \times 10^{-4} c$. Note that $v_2$ vanishes 
at the polar axis and at the equator, so that the average velocity is 
substantially smaller than that shown in Fig.~\ref{fig7.8}. Taking into account 
that the total mass in the pre-shock region is of the order of $M_\mathrm{ps}
\sim 1 M_{\odot}$, the kinetic energy of the oscillations
is roughly
\begin{equation}
E_\mathrm{kin} \approx \frac{1}{2} M_\mathrm{ps} (\hat{v}_2)^2 \approx 2 \times
10^{-8} M_{\odot} c^{2}.
\end{equation}
This energy is comparable to the total energy radiated in gravitational 
waves in a typical core collapse event~\cite{ZwM97,DFM022}.
Transferring such an amount of energy to the pre-shock region 
seems unphysical, as gravitational waves interact with matter 
only very weakly. Instead, as we describe next, we conclude 
that the oscillations are mainly introduced by our choice of 
coordinates.

Following the work of Bishop et al.~\cite{BGL97} inertial coordinates
can be established at future null infinity $\Scri^{+}$. The angular
inertial coordinate $\theta_B$ can be constructed solving the partial
differential equation
\begin{equation}
\label{inerttheta}
(\partial_u + U \partial_{\theta}) \theta_B = 0,
\end{equation}
with initial data $\theta_B(u=0) = \theta(u=0)$. Instead of
solving Eq.~(\ref{inerttheta}) directly,
we determine its characteristic curves,
\begin{eqnarray}
\frac{d \theta}{d u} & = & U (\theta,u),\\
\theta(u=0) & = & \theta_B,
\end{eqnarray}
along which $\theta_B$ is constant. With suitable interpolations,
$\theta_B$ can then be determined for arbitrary angles $\theta$.

Making use of Eq.~(\ref{inerttheta}), it is possible to define
an ``inertial'' meridional 4-velocity component
\begin{equation}
\label{uthetaB}
u^{\theta_B} = \frac{\partial \theta_B}{\partial \theta} \Big|_{u}^{\footnotesize{\Scri^+}}
(u^{\theta} - U \Big|^{\footnotesize{\Scri^+}} u^u).
\end{equation}
The dashed line in Fig.~\ref{fig7.8} shows the corrected (``inertial'')
meridional velocity $r u^{\theta_B}$. Remarkably, the oscillations have
almost disappeared, which clearly shows that gauge effects can play
a major role for the collapse dynamics in the pre-shock region.

\section{Gravitational waves}
\label{sec:GWfromSN}

\subsection{Quadrupole gravitational waves}

The common approach to the description of gravitational waves for
a fluid system relies on the quadrupole formula~\cite{LaL61}. The standard
quadrupole formula is valid for weak sources of gravitational waves under the 
assumptions of slow motion and wave lengths of the emitted gravitational 
waves smaller than the typical extension of the source. The requirement 
that the sources of gravitational waves are weak includes the requirement 
that the gravitational forces inside the source can be neglected. This first 
approximation can be extended based on Post-Newtonian expansions (for a 
detailed description see the recent review~\cite{Bla02} and references therein).

In a series of papers~\cite{Win83,Win84,IWW84,Win87}, Winicour established 
that the quadrupole radiation formula can be derived in the Newtonian
limit of the characteristic field equations. Let $Q$ be the quadrupole
moment transverse to the $(\theta, \phi)$ direction
\begin{equation}
\label{defq}
Q = q^A q^B \Big(\frac{x^i}{r} \Big)_{,A} \Big(\frac{x^j}{r} \Big)_{,B} Q_{ij},
\end{equation}
where 
\begin{equation}
Q_{ij} = \int \rho (x_i x_j - \delta_{ij} r^2/3) d^3x
\end{equation}
is the quadrupole tensor and $q_{A}$, $A=2,3$, is the complex dyad for 
the unit sphere metric
\begin{equation}
d\theta^2 + \sin^2 \theta d\phi^2 = 2 q_{(A} q_{B)} dx^A dx^B.
\end{equation}
As usual we use parentheses to denote the symmetric part.
For our axisymmetric setup, Eq.~(\ref{defq}) reduces to
\begin{equation}
\label{quadmom}
Q = \pi \sin^2 \theta \int_{0}^{R} dr' \int_{0}^{\pi} \sin \theta' d \theta'
r'^{4} \rho \Big( \frac{3}{2} \cos^2 \theta' - \frac{1}{2} \Big).
\end{equation}
On the level of the quadrupole approximation~\cite{Win87} the {\it quadrupole news} 
$N_{0}$ reads
 \begin{equation}
\label{quadnews}
N_{0} = \frac{d^3}{du_B^3} Q.         
\end{equation}
With our null foliation it is natural to evaluate the quadrupole
moment~(\ref{quadmom}) as a function of retarded time, i.e., for the evaluation
of the integral we completely relax the assumption of slow motion. 

It is well known~\cite{Fin89} that the third numerical time derivative 
appearing in Eq.~(\ref{quadnews}) can lead to severe numerical problems 
resulting in numerical noise which dominates the quadrupole signal. 
Therefore, we make use of the fluid equations in the Newtonian limit to 
eliminate one time derivative. Defining the ``Newtonian velocities"
\begin{eqnarray}
v_1 & = & u^r = \frac{dr}{dx} u^x, \\
v_2 & = & r u^{\theta} = r \frac{u^y}{\sin \theta},
\end{eqnarray}
the quadrupole radiation formula~(\ref{quadnews}) can be rewritten with the use
of the continuity equation as the so-called {\it first moment of momentum
  formula }
\begin{eqnarray}
\label{fmomf}
N_0 &=& \frac{d^2}{du_B^2} \Big( \pi \sin^2 \theta \int_{0}^{R} dr'
\int_{0}^{\pi} \sin \theta' d \theta' r'^{3} 
\nonumber
\\
&\times&
\rho ( v_1 (3 \cos^2 \theta' - 1) - 3 v_2 \sin \theta' \cos \theta' ) \Big). 
\end{eqnarray}
We henceforth work with Eqs.~(\ref{quadnews}) and~(\ref{fmomf}) for
estimating the quadrupole radiation. In addition, following earlier
work~\cite{MSM91,ZwM97}, we define the quantity $A_{20}^{E2}$, which
enters the total power radiated in gravitational waves in the
quadrupole approximation as
\begin{equation}
\label{quadpower}
\frac{dE}{du_B} = \frac{1}{32 \pi} \Big( \frac{dA_{20}^{E2}}{du_B} \Big)^2.
\end{equation}
$A_{20}^{E2}$ also 
arises as coefficient for the quadrupolar term in the expansion
of the quadrupole strain (i.e. the gravitational wave signal) 
$h_+$ in spherical harmonics~\cite{note2}
\begin{equation}
h_+(u_B) = \frac{1}{8} \sqrt{\frac{15}{\pi}} \sin^2 \theta \frac{A_{20}^{E2}(u_B)}{R},
\end{equation}
where $R$ denotes the distance between the observer and the source.
$A_{20}^{E2}$ can be deduced from the quadrupole moment as
\begin{eqnarray}
\label{A20}
\nonumber
A_{20}^{E2} &=& \frac{16}{\sqrt{15}} \pi^{\frac{3}{2}}
\frac{d^2}{du_B^2} \Big[ \int_{0}^{R} dr' \\
& & \int_{0}^{\pi} \sin \theta' d \theta'
r'^{4} \rho \Big( \frac{3}{2} \cos^2 \theta' - \frac{1}{2} \Big) \Big],
\end{eqnarray}
or alternatively using the first moment of momentum formula in order to 
eliminate one time derivate, in analogy to the transition from Eq.~(\ref{quadnews}) to 
Eq.~(\ref{fmomf}), i.e.,
\begin{eqnarray}
\label{a20fmom}
A_{20}^{E2} &=& \frac{16}{\sqrt{15}} \pi^{\frac{3}{2}}
\frac{d}{du_B} \Big[ \int_{0}^{R} dr'
\int_{0}^{\pi} \sin \theta' d \theta' r'^{3} 
\nonumber
\\
&\times&
\rho ( v_1 (3 \cos^2 \theta' - 1) - 3 v_2 \sin \theta' \cos
\theta' ) \Big].
\end{eqnarray}

As shown in Fig.~\ref{fig7.9} we find good agreement when computing the wave 
strain $A_{20}^{E2}$ using Eq.~({\ref{A20}) and Eq.~(\ref{a20fmom}). In order 
not to have the time derivatives dominated by numerical noise, we have averaged 
the matter contribution in the integrands of Eq.~(\ref{A20}) and 
Eq.~(\ref{a20fmom}) over a few neighboring grid points before calculating the 
time derivatives.

\begin{figure}[t]
\centerline{\psfig{file=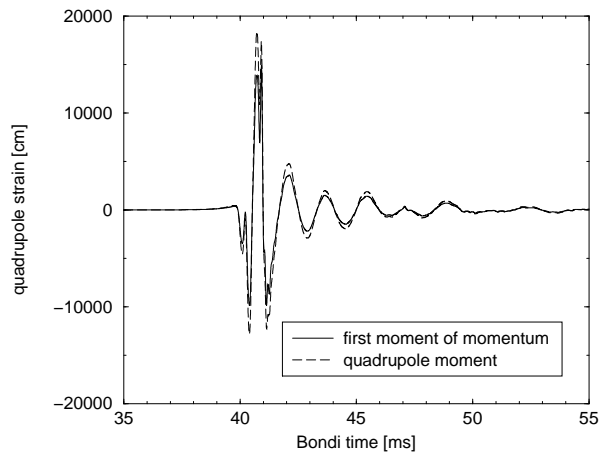,width=0.7 \linewidth,angle=-90}}
\caption{Gravitational wave strain $A_{20}^{E2}$ for the simulation of
the collapse model $\mathfrak{B}01$. The solid curve shows the result
using the first moment of momentum approach Eq.~(\ref{a20fmom}), the dashed line
is based on Eq.~(\ref{A20}). The good agreement found between both
approaches shows that our general relativistic fluid evolution is
internally consistent.}
\label{fig7.9}
\end{figure}

This result checks the implementation of the continuity equation and,
as this equation is not calculated separately but as a part of a system 
of balance laws, it also checks the overall implementation of the fluid 
equations in the code. We note that the equivalence between Eq.~(\ref{A20})
and Eq.~(\ref{a20fmom}) is only strictly valid in the Minkowskian
limit and for small velocities, which is the origin for the observed small
differences between the curves in Fig.~\ref{fig7.9}. Substituting $\rho$ 
by $\rho u^u e^{2 \beta}$ in Eq.~(\ref{A20}) and by $\rho e^{2 \beta}$ 
in Eq.~(\ref{a20fmom}), by which we restore the equivalence in a general 
relativistic spacetime, we find excellent agreement between the two 
approaches for calculating $A_{20}^{E2}$.

\begin{figure}[t]
\centerline{\psfig{file=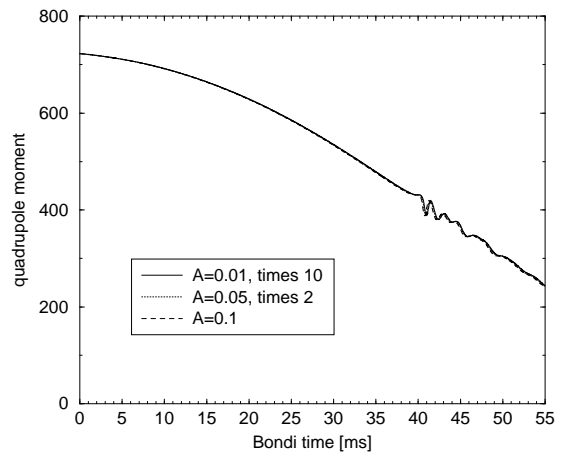, width=0.7 \linewidth,angle=-90}}
\caption{Quadrupole moment Q (in
units $c=G=M_{\odot}=1$) as a function of time for three models of
type $\mathfrak{B}$ with
perturbation amplitude $A=0.01$, $A=0.05$ and $A=0.1$. The first two
results are rescaled with respect to $A=0.1$ assuming a linear
dependence. All three curves overlap in the diagram. The
quadrupole moment (and hence the quadrupole signal) scales linearly
with the amplitude of the perturbation in the chosen parameter
region.}
\label{fig7.10}
\end{figure}

Since we are imposing only small perturbations from spherical symmetry, we
expect a linear dependence of the non-spherical dynamics and the
gravitational wave signal as a function of the perturbation amplitude.
We have verified in a series of runs that the amplitude of the quadrupole 
moment (and thus the quadrupole radiation signal) indeed scales linearly 
with the amplitude of the initial perturbations (see Fig.~\ref{fig7.10}). 
This observation marks another important test for the correctness of the 
global dynamics of our code.

On the other hand, when comparing the quadrupole news defined in 
Eq.~(\ref{quadnews}) or  Eq.~(\ref{fmomf}) with the Bondi news signal $N$ 
evaluated at future null infinity (which is defined in Eq.~(\ref{bondinews})
below), we find important discrepancies. This can be seen in Fig.~\ref{fig7.11}, 
where we plot both, the Bondi news and the quadrupole news for model
$\mathfrak{B}01$. We note that the differences manifest themselves
not only in the amplitude of the oscillations, but also in the 
frequencies of the signals. This behavior is clearly different 
from the one we observed in the studies of neutron star pulsation 
carried out in Ref.~\cite{SFM02}, where both signals showed very
good agreement.

\begin{figure}[t]
\centerline{\psfig{file=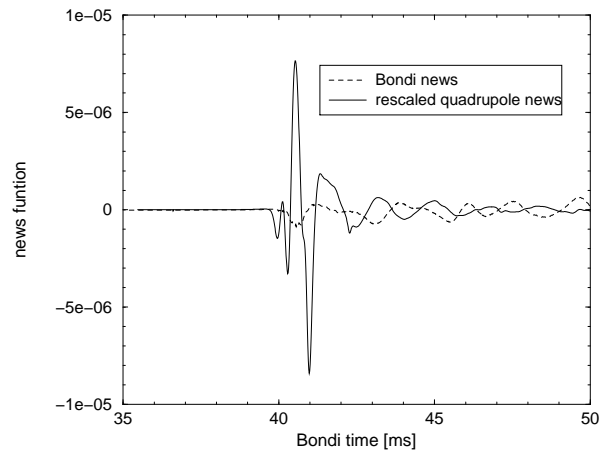,width=0.7 \linewidth,angle=-90}}
\caption{Bondi and quadrupole news as a function of time for model
$\mathfrak{B}01$. The solid curve corresponds to the quadrupole 
news according to Eq.~(\ref{fmomf}), the dashed curve to the Bondi news 
signal. For visualization reasons, we have divided the quadrupole news 
result by 50. Remarkable disagreement is found between both signals.}
\label{fig7.11}
\end{figure}

As mentioned above, the quadrupole formula is only the first term in a
Post-Newtonian expansion for the gravitational radiation. The next,
non-vanishing contribution to the gravitational strain for our axisymmetric 
configuration is the hexadecapole contribution,
which reads~\cite{MSM91}
\begin{equation}
\label{hexadecapole}
h_+^{HD} = \frac{9}{8}\sqrt{\frac{5}{\pi}} \sin^2 \theta (1 - \frac{7}{6} \sin^2
\theta) \frac{A_{40}^{E2}}{R}.
\end{equation}
The quantity $A_{40}^{E2}$ is defined as
\begin{eqnarray}
A_{40}^{E2} & = & \frac{d^4}{du_B^4} M_{40}^{E2}, \\
\label{hexadecamom}
M_{40}^{E2} & = & \frac{\sqrt{5}}{63} \pi^{\frac{3}{2}} \int_{0}^{R} dr'
\int_{0}^{\pi} \sin \theta' d \theta' r'^{6} 
\nonumber
\\
&\times&
\rho ( 7  \cos^4 \theta' - 6 \cos^2 \theta' + \frac{3}{5}),
\end{eqnarray}
or alternatively
\begin{eqnarray}
A_{40}^{E2} & = & \frac{d^3}{du_B^3} N_{40}^{E2}, \\
\nonumber
N_{40}^{E2} & = & \frac{4 \sqrt{5}}{63} \pi^{\frac{3}{2}} \int_{0}^{R} dr'
\int_{0}^{\pi} \sin \theta' d \theta' r'^{5} 
\nonumber
\\ \nonumber
&\times&
\rho \Big(v_1 ( 7  \cos^4 \theta' - 6 \cos^2 \theta' + \frac{3}{5}) \\
\label{hmomf}
 &    & + v_2 (3 - 7 \cos^2 \theta') \sin \theta' \cos \theta' \Big)  .
\end{eqnarray}

By extracting the hexadecapole moment for the above result,
we found, however, that the associated amplitude is too small in order
to explain the observed differences in
Fig.~\ref{fig7.11}. In addition, one would expect in general
 that the contribution of the hexadecapole moment increases the
amplitude of the approximate signal. However, the
amplitude of the quadrupole news in Fig.~\ref{fig7.11} is already much 
{\it larger} than that of the Bondi news evaluated at $\Scri^{+}$.

As we discussed in the preceding section, the global dynamics of the core 
collapse and bounce is correctly reproduced with our numerical code
(see also the validation tests in the Appendix). We have strong evidence 
that the quadrupole signals extracted from our collapse simulations do not 
correspond to physical gravitational wave signals. In the following, we 
describe the different arguments which support this claim.

First, if the quadrupole radiation signal corresponded to the true
physical signal, it would be very difficult to understand why the Bondi 
signal has a significantly smaller amplitude. In the calculation of the 
Bondi news, Eq.~(\ref{bondinews}), the contribution of the different terms are
relatively large and add up to a small signal (see below). Under the 
assumption that the quadrupole news signal is correct and the Bondi 
news signal is wrong, it is extremely unlikely that possible 
errors in the contribution to the Bondi news add up to a very small signal.

Second, we have performed comparisons between our numerical code and
the code of Refs.~\cite{DFM02,DFM022}, finding much 
larger amplitudes for the quadrupole gravitational wave signal in our 
case. However, we note that comparing the results of both codes in
axisymmetry is ambiguous, as possible differences might have different
explanations. For example, the use of the conformally flat metric
approach in~\cite{DFM02,DFM022} is clearly an approximation 
to general relativity, which should create some differences. Furthermore, 
the coordinate systems used in both codes for the
computation of the quadrupole moment are different. Only in our code,
the quadrupole moment is evaluated on a light cone, i.e. as a function
of retarded time.

\begin{figure}[t]
\centerline{\psfig{file=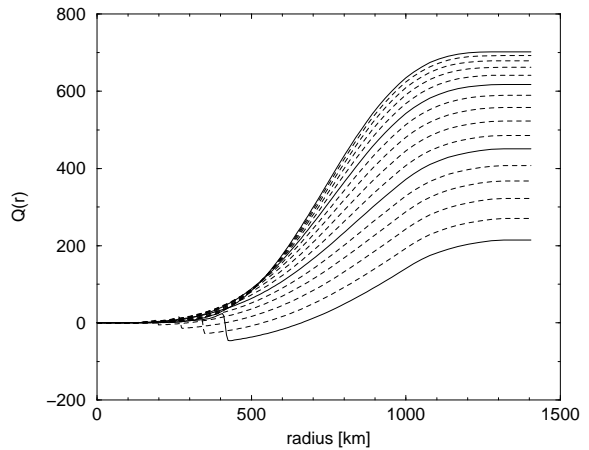,width=0.7 \linewidth,angle=-90}}
\caption{Radial contribution to the quadrupole moment. We plot the value of the 
integral $Q(r) = \pi \sin^2 \theta \int_{0}^{r} dr' \int_{0}^{\pi} \sin \theta' 
d \theta' r'^{4} \rho ( \frac{3}{2} \cos^2 \theta' - \frac{1}{2} )$ as a function 
of the radial coordinate $r$ for different values of time. The data is plotted 
after a fixed number of time steps, starting with initial data at
$u_B=0$ ms (upper solid curve).
The data was taken from a simulation of model
$\mathfrak{B}01$. Large amplitude oscillations of the quadrupole moment, as 
they can be seen in Fig.~\ref{fig7.10}, can only be created - at least shortly 
after bounce - in the outer region of the infalling matter well in front of the 
shock.}
\label{fig7.12}
\end{figure}

A third and physically motivated argument stems from the spatial distribution 
of matter in our simulation. As it can be seen from Fig.~\ref{fig7.12}, the main 
contribution to the radial integral of the quadrupole moment comes from the
outer, infalling layers of matter. These outer layers are responsible for the 
oscillations in the quadrupole moment, which can be seen in Fig.~\ref{fig7.10}.
Following the same reasoning as in the previous section it is obvious to conclude
that the calculation of the quadrupole moment is also affected by
our choice of coordinates, i.e. by gauge effects.

For all these reasons we extract the quadrupole moment in the angular
coordinate system defined by Eq.~(\ref{inerttheta}). However, introducing 
the inertial angular coordinate does not help to obtain a better agreement 
between quadrupole and Bondi signals, the extracted quadrupole moment 
almost agrees with the results shown in Fig.~\ref{fig7.10}. Since the 
difference of Bondi time between the different angular directions on 
our Tamburino-Winicour foliation is in general of the same order as the 
lapse of time for one time step, we expect a similar result when evaluating 
the quadrupole moment at a fixed inertial time. However, by prescribing 
the necessary coordinate transformations to define Bondi coordinates only 
at $\Scri^+$, we do not take into account an inertial radial coordinate,
which should be used for the evaluation of the quadrupole moment.

As already mentioned before, in Ref.~\cite{SFM02} we found good agreement 
between the Bondi signal and the quadrupole signal when calculating gravitational 
waves from pulsating relativistic stars. Hence, the obvious question
arises of why the  quadrupole formula could be applied in those scenarios. The answer lies
in the small velocities encountered in the problem of neutron star pulsations.
Whereas the typical maximum fluid velocities in the oscillation problem are of 
the order of $10^{-5}c-10^{-4}c$, fluid velocities of up to $0.2c$ are reached
for the core collapse scenario. Furthermore, due to the non-spherical
dynamics of the proto-neutron star formed in the interior of the collapsed 
region, the metric can pick up gauge contributions which are created as a
consequence of our requirement to prescribe a local Minkowski frame at
the vertex of the light cones. Gauge contributions may also play a more 
important role in the collapse scenario due to the enlarged radial 
extension of the collapsing iron core (about 1500 km), which is much 
larger than the corresponding one for neutron star pulsations (about 15 km).

We note that since the collapse involves fluid velocities of up to
$0.2 c$, it is not obvious whether the functional form for the quadrupole 
moment established in the slow motion limit on the light cone will still 
be valid. In fact, the situation could be similar to the case of the
total mass of spacetime, where a naive definition, even in spherical
symmetry, as
\begin{equation}
M_n = 4 \pi \int_0^R r'^2 \rho (1 + \epsilon) dr',
\end{equation}
would only be a valid approximation for small fluid velocities. This can 
be understood from the comparison with the expression of the Bondi mass
in the form
\begin{equation}
M_B = 4 \pi \int_0^R r'^2 [\rho (1 + \epsilon) (- u^u u_u) - p (1
+ u^u u_u)] dr',
\end{equation}
(no summation is involved in this expression). Only vanishing fluid 
velocities, i.e. $u^u u_u = -1$, ensure that the two masses are equal, 
$M_n = M_B$.

We experimented with possible alternative functional forms for the 
quadrupole moment which result in significant differences. 
An unambiguous clarification 
of which functional form has to be used for the quadrupole moment in 
the extended regime of validity of large fluid velocities could only 
be obtained by a derivation of the quadrupole formula in the Tamburino 
gauge. However, technical complications for such a derivation are so 
severe that it has only been accomplished for a simplified radiating 
dust model~\cite{IWW83} (see the related discussion in Ref.~\cite{Win87}).

\subsection{The Bondi news signal}

The numerical extraction of the Bondi news is a very complicated undertaking.
Reasons for possible numerical problems are diverse: First, its extraction 
involves calculating {\it non-leading} terms from the metric expansion at 
future null infinity. All 
the metric quantities are global quantities, and are thus sensitive to any
numerical problem in the entire computational domain. Second, when
calculating the gravitational signal in the Tamburino-Winicour
approach, one has to take into account gauge effects. For the
present calculations of the gravitational wave signal from
core collapse, the gauge contributions are indeed the {\it dominant}
contribution, which can easily influence the physical signal.

We have described in detail the formalism and numerical methods to deal 
with gravitational waves without approximation in our axisymmetric
characteristic code in Ref.~\cite{SFM02}. In the following, we will only 
repeat the most important aspects. The total energy emitted by gravitational
waves to infinity during the time interval $[u,u+du]$ in the angular direction
$[y, y + d y]$ is given by the expression
\begin{equation}
dE = \frac{1}{2} N^2 \omega^3 e^{2H} dy du,
\end{equation}
where the Bondi news function $N$ reads
\begin{eqnarray}
\label{bondinews}
N &=& \frac{1}{2} \frac{e^{-2H}} {\omega^{2}}
\left\{2 c_{,u} + \frac{(\sin\theta
\ c^{2} \ L),_\theta}{\sin\theta \ c} 
\right .
\nonumber \\
& & \left . + e^{-2K} \omega \sin \theta [\frac{(e^{2H}
\omega)_{, \theta}}{\omega^{2} \sin \theta}]_{, \theta} \right\}.
\end{eqnarray}
$K, c, H$ and $L$ are defined by a power series
expansion of the metric quantities at $\Scri^{+}$ as follows,
\begin{eqnarray}
\gamma & = & K + \frac{c}{r} + O(r^{-2}),\\
\beta & = & H + O(r^{-2}),\\
U & = & L + O(r^{-1}).
\end{eqnarray}

\begin{figure}[t]
\centerline{\psfig{file=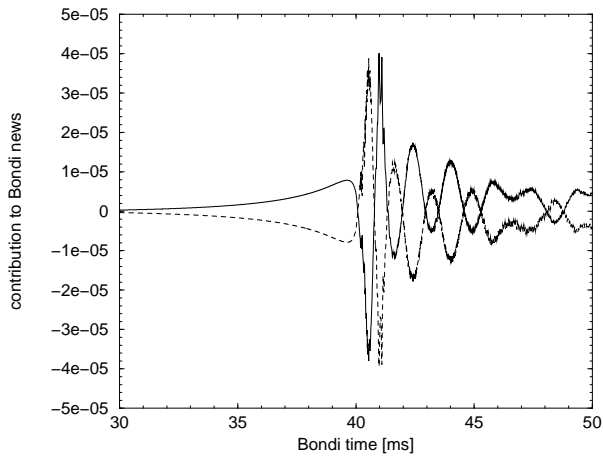,width=2.4in,angle=-90}}
\caption{Different contributions to the Bondi news. The solid curve
corresponds to the term involving $c_u$ (first addend) in
Eq.~(\ref{bondinews}), the dashed curve to the contribution from the second
and third addend. By summing up both contributions we obtain the Bondi
news, which is close to zero. In addition, we note that when separating
the third addend into angular derivatives of $H$ and $\omega$, each single 
contribution has an amplitude 23 times larger than what is shown in the 
figure.}
\label{fig7.14}
\end{figure}

We plot in Fig.~\ref{fig7.14} the different
contributions to the Bondi news for the collapse model
$\mathfrak{B}01$. It becomes clear from this plot that a very
accurate determination of the metric fields is essential.
As it can be further seen in this figure, the metric quantities show high frequency 
numerical noise, as soon as the shock forms (at a Bondi time of 
about $40$ ms). In order to demonstrate that the noise is actually created 
at the shock, we plot in Fig.~\ref{fig7.15} the location of the shock together 
with the gravitational wave signal. Clearly, the noise is created by the 
motion of the shock across the grid, its temporal behavior following the 
discontinuous jumps of the shock between adjacent grid cells. We note that 
due to the coarser radial resolution used in the outer layers of the core, 
the frequency of the noise slowly decreases with time.

\begin{figure}[t]
\centerline{\psfig{file=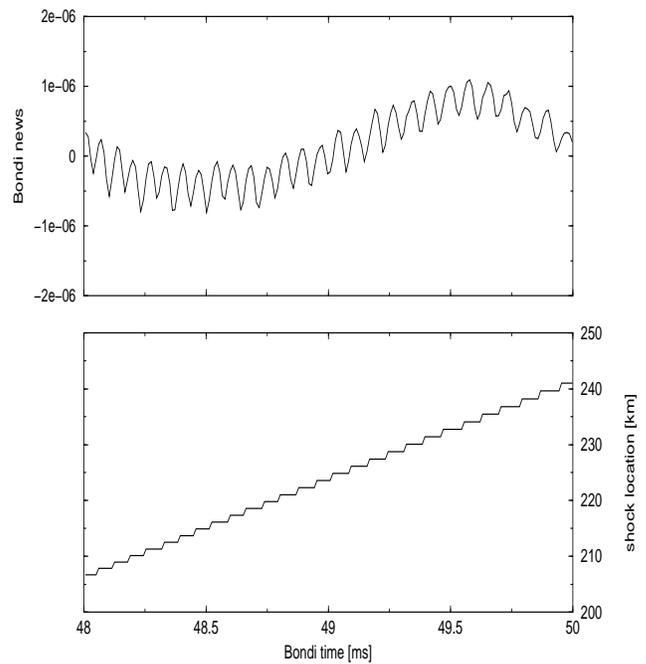,width=3.5in,height=3.2in,angle=-90}}
\caption{Upper panel: Bondi news as a function of time. High frequency
noise is overlayed on top of a small frequency modulation. Lower panel: 
Time evolution of the radial location of the cross section of the shock 
front with the equator. Due to the finite resolution, the location of the 
shock wave moves discontinuously. The frequency of these jumps coincides
with that of the noise in the Bondi news. Once created at the shock, the
noise is propagated instantaneously to infinity through the numerical solution 
of the metric equations.}
\label{fig7.15}
\end{figure}

As we have pointed out in the previous section, the shock front is
well captured in only a few radial zones with our high-resolution
shock-capturing scheme. It might seem surprising that a small localized
error created in a few radial zones can have such a large effect on the
gravitational wave signal. However, one has to keep in mind that the
radial integration of the metric variables picks up this error and
propagates it to future null infinity instantaneously. It is important
to stress that the effect of the numerical noise on the dynamics of 
the collapse and bounce is entirely negligible. However, the extraction
of the Bondi news signal is extremely sensitive to it.

We have verified that the frequency of the noise increases, as expected, 
with radial resolution. Unfortunately, its amplitude does not decrease
substantially with radial resolution, at least not in the resolution 
regime accessible to us~\cite{note3}. Therefore, we tried to eliminate 
the noise by different methods. In a first attempt, we smoothed out
the shock front, either in the hydrodynamical evolution itself or
before using the fluid variables in the source terms of the
metric equations. In both cases, it was impossible to obtain a
smooth signal without changing the dynamics. In a second attempt, 
following the work of~\cite{Gom01}, we rearranged the metric equations 
eliminating second derivatives which might be ill-behaved at the shock. 
Defining a metric quantity
\begin{equation}
X = r^2 f^2 e^{2(\gamma-\beta)} \hat{U}_{,x} - 2 (\beta_{,y} - (1-y^2) 
\hat{\gamma}_{,y}),
\end{equation}
and solving the hypersurface equations successively for $\beta$, $X$, $\hat{U}$ 
and $S$, it is possible to eliminate all second derivatives from the 
hypersurface equations. Unfortunately, the noise is not significantly 
reduced by this rearrangement of the metric equations. Finally, going 
to larger time steps for the fluid evolution only - solving the metric 
equations several times between one fluid time step - was not effective 
either.

\begin{figure}[t]
\centerline{\psfig{file=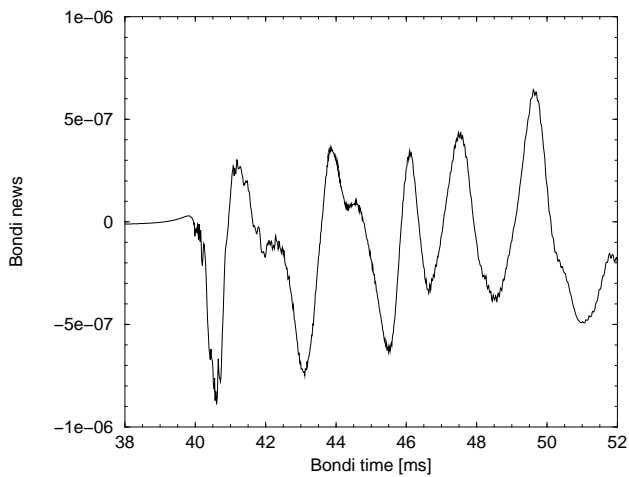,width=2.5in,angle=-90}}
\caption{Bondi news as a function of Bondi time for the collapse model
$\mathfrak{B}01$. The displayed time interval covers the late collapse
stage until several ms after bounce at about $t=40$ ms. 
During the collapse stage, the 
gravitational wave signal is negligible. After bounce a complicated 
series of oscillations sets in.
}
\label{fig7.16}
\end{figure}

After these attempts we decided to eliminate the noise from the gravitational 
wave signals only after the numerical evolution. We experimented with two 
different smoothing methods. In the first method, we calculate the Fourier 
transform of the data, and eliminate all frequencies beyond a certain threshold
frequency (of about 5 - 10 kHz). Then, when transforming back from Fourier space
all the high-frequency part of the data is removed. In a second method we 
simply average the signal over a few neighboring points. We have applied 
this second method in what is described below.

Fig.~\ref{fig7.16} shows the Bondi news signal for the collapse
model $\mathfrak{B}01$. The figure focuses on the part of the signal
around bounce. After the initial gravitational wave content is
radiated away (in the first $5$ ms, not depicted in the figure), the signal
in the collapse stage is very weak. This is expected, as the dynamics
is well reproduced by a spherical collapse model during this stage. At
bounce, the Bondi news shows a spike. Afterwards, a complicated 
series of oscillations is created due to the pulsations of the forming 
neutron star and the outward propagation of the shock. Typical 
oscillation frequencies are of the order of $0.5 - 1$ kHz, at which 
the current gravitational wave laser interferometers have maximum 
sensitivity. 

Correspondingly, Fig.~\ref{fig7.17} shows the Bondi news signal for
the collapse model $\mathfrak{C}01$. Here again, after radiating away
the initial gravitational wave content, the collapse phase is
characterized by very small radiation of gravitational waves. At
bounce, we again observe a strong spike in the signal. Afterwards,
the oscillations in the signal are rather rapidly damped.

\begin{figure}[t]
\centerline{\psfig{file=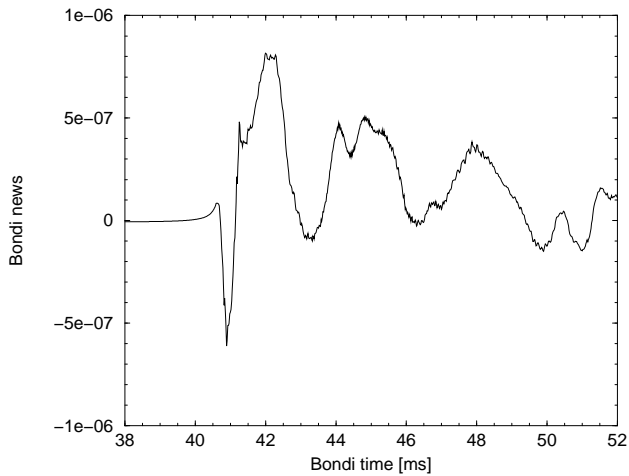,width=2.5in,angle=-90}}
\caption{Bondi news as a function of time for the collapse model
$\mathfrak{C}01$. The bounce at about 41 ms is characterized by
a large spike in the gravitational wave signal. After bounce,
the signal shows oscillations, with a principal frequency of
about $0.35$ kHz.
}
\label{fig7.17}
\end{figure}

We stress that as a consequence of the necessary smoothing techniques 
applied, only the main features of the gravitational wave signals in 
Figs.~\ref{fig7.16} and~\ref{fig7.17} are reliably reproduced. This 
also applies to possible offsets of the Bondi news, which affect in 
particular the gravitational wave strain. 
Comparing the Bondi news function for the different collapse models of 
type $\mathfrak{B}$, we observe to good approximation a linear 
dependence of the Bondi news with the perturbation amplitude. This is 
reflected in the total energy radiated away in gravitational waves, 
which scales quadratically with the amplitude of the initial perturbation. 
A summary of the results on the gravitational wave energy is listed in
in Table~\ref{tab:bradenergy}.

\begin{table}[h!]
\centering
\caption{Total energy radiated in gravitational waves during the first
50 ms for the collapse simulations of type $\mathfrak{B}$. The
initial gravitational wave content is the dominant contribution to the
total energy. This energy scales quadratically with the amplitude
of the initial perturbation, as can be inferred from the last column,
where the corresponding energies have been rescaled with respect
to that of collapse model $\mathfrak{B}01$.}
\medskip
\begin{tabular}{|c|c|c|}
\hline
model  & total energy radiated [$M_{\odot}$] & rescaled result
[$M_{\odot}$] \\
\hline \hline
$\mathfrak{B}001$  &  $4.31 \times 10^{-9}$ & $4.31 \times 10^{-7}$
\\
$\mathfrak{B}005$  &  $1.08 \times 10^{-7}$ & $4.32 \times 10^{-7}$
\\
$\mathfrak{B}01$   &  $4.32 \times 10^{-7}$ & $4.32 \times 10^{-7}$
\\
\hline
\end{tabular}
\vspace{0.5cm}
\label{tab:bradenergy}
\end{table}

\section{Discussion}
\label{sec:discussion}

We have presented first results from axisymmetric core collapse simulations in
general relativity. Contrary to traditional approaches, our framework uses a 
foliation based on a family of light cones, emanating from a regular center, 
and terminating at future null infinity. To the best of our knowledge, the 
characteristic formulation of general relativity has never been used
before in simulations of supernova core collapse 
and in the extraction of the associated exact gravitational waves. 
Our axisymmetric hydrodynamics code is accurate enough to allow for a detailed 
analysis of the global dynamics of core collapse in general. But we have not found 
a robust method for the (Bondi news) gravitational wave extraction in the 
presence of strong shock waves.

Comparing our results to other recent work on relativistic supernova
core collapse~\cite{DFM01,DFM022}, it is not surprising that 
numerical noise in the gravitational waveforms is more noticeable in our 
approach. Whereas in the conformal flat metric approach employed in 
\cite{DFM01,DFM022} the metric equations of general relativity 
reduce to elliptic equations, which naturally smooth out high-frequency numerical 
noise, we solve for the gravitational wave degrees of freedom directly using the 
full set of field equations of general relativity, and hence we have to solve a
hyperbolic equation. It remains to be seen whether a similar numerical noise
to the one we find when extracting the gravitational wave signal will be
encountered in core collapse simulations solving the full set of Einstein
equations in the Cauchy approach. In this respect we mention recent
axisymmetric simulations by Shibata using a conformal-traceless reformulation
of the ADM system~\cite{Shi02} where, despite of the fact that long-term 
rotational collapse simulations could be accurately performed, gravitational
waves could not be extracted from the raw numerical data since
their amplitude is much smaller than that of 
other components contained in the metric and/or numerical noise.

With the current analysis we have presented in this paper, it is not obvious
how the numerical noise of the Bondi news can be effectively eliminated.
Including rotation in the simulations, which would be the natural next step
for a more realistic description of the scenario, could help in this respect.
Due to the global asphericities introduced by rotation, one would expect,
in general, gravitational wave signals of larger amplitude, which could make
the numerical noise less important, if not completely irrelevant. In
addition to this possibility we propose the following methods to improve the
extraction of the
gravitational wave signals: In a first approach one should try to rearrange
the metric equations by introducing auxiliary fields which could effectively
help to diminish the importance of high-order derivatives, especially of the
fluid variables, which can be discontinuous. Unfortunately, to the best of our
knowledge, there is no clear guideline to what is really needed to eliminate
the numerical noise completely, apart from the hints given by~\cite{Gom01}. Our
attempts in this direction have not yet been successful, but we believe there
is still room for improvement. Alternatively, one should try to implement
pseudospectral methods for the metric update. Pseudospectral methods would
allow for a more efficient and accurate numerical solution of the metric
equations. In a third promising line of research we propose to consider the
inclusion of adaptive grids and methods of shock fitting into the current code. 
With the help of an adaptive grid, one could try to arrange the entire core 
collapse simulation in such a way that the shock front always stays at a fixed
location of the numerical grid. By avoiding the motion of the shock front 
across the grid, one would expect the noise in the gravitational wave signals 
to disappear. But already increasing the radial resolution substantially in the 
neighborhood of the shock front could help to obtain an improved representation 
of the shock. All these issues are ripe for upcoming investigations.

\section{Acknowledgements}

It is a pleasure to thank Harald Dimmelmeier for helpful discussions and
for performing reference runs with his numerical code. We further
thank Masaru Shibata for comments. Our work has been 
supported in part by the EU Programme 'Improving the Human Research Potential 
and the Socio-Economic Knowledge Base', (Research Training Network Contract
HPRN-CT-2000-00137). P.P. acknowledges support from the Nuffield
Foundation (award NAL/00405/G). J.A.F acknowledges support from a
Marie Curie fellowship from the European Union (HPMF-CT-2001-01172)
and from the Spanish Ministerio de Ciencia y Tecnolog\'{\i}a
(grant AYA 2001-3490-C02-01).


\appendix

\section{}
\label{appendix}


In this Appendix we present tests specifically aimed to calibrate our code
in core collapse simulations. The reader is addressed to Ref.~\cite{SFM02}
for information on further tests the code has successfully passed concerning,
among others, long-term evolutions of relativistic stars and mode-frequency 
calculations of pulsating relativistic stars. 

\subsection{Shock reflection test}

In order to assess the shock-capturing properties of the code, we have 
performed a shock reflection test in Minkowski spacetime. This is a standard 
problem to calibrate hydrodynamical codes~\cite{MaM99}. 
A cold, relativistically 
inflowing ideal gas is reflected at the origin of the coordinate system, 
which causes the formation of a strong shock. We start the simulation with 
a constant density region, where $\rho=\rho_0$, $u^r=u^r_\mathrm{R}$ and 
$\epsilon=\epsilon_\mathrm{R} = 0$ (we set $\epsilon \approx 10^{-11}$ for 
numerical reasons). From the continuity equation it follows that the rest 
mass density in the unshocked region obeys
\begin{equation}
\rho_\mathrm{R}(u,r) = \rho_0 \left( 1 - \frac{u^r_\mathrm{R}}
{u^u_\mathrm{R} r} u \right)^2.
\end{equation}
From momentum conservation arguments, it is clear that the velocity in the
shocked region vanishes, $u^r_\mathrm{L}=0$. Evaluating the Rankine-Hugoniot
jump conditions for the fluid equations, we obtain:
\begin{eqnarray}
s & = & \frac{(\Gamma - 1) \epsilon_\mathrm{L}}
              {u^u_\mathrm{R} -1 - \Gamma \epsilon_\mathrm{L}}, \\ 
\epsilon_\mathrm{L} & = & u^u_\mathrm{R} + u^r_\mathrm{R} - 1, \\
\rho_\mathrm{L} & = & \rho_s \frac{\Gamma (u^r_\mathrm{R})^2 - 
(\Gamma - 1) \epsilon_\mathrm{L}}
                         {(\Gamma -1) \epsilon_\mathrm{L}}, \\
p_\mathrm{L} & = & (\Gamma - 1) \rho_\mathrm{L} \epsilon_\mathrm{L}.
\end{eqnarray}
Here, $s$ denotes the shock speed and $\rho_s = \rho_\mathrm{R}(u,r = su)$ the
rest mass density in front of the shock.

We performed this test with different values of the fluid velocity, and
different schemes for the fluid evolution. Fig.~\ref{fig4.10} shows the
results for an ultrarelativistic flow ($u^{r} = -0.9999c$).
For this particular test we used the HLL solver and increased the
numerical viscosity by a factor 2 in order to damp small post-shock
oscillations. The agreement with the analytic solution is satisfactory, and
the shock front is very steep, being resolved with only one or two radial 
zones. The deviation close to the origin is a well-known failure of 
finite-difference schemes for this problem (see, e.g.~\cite{Noh87}), 
which is not important for our purposes.

\begin{figure}[t]
\centerline{\psfig{file=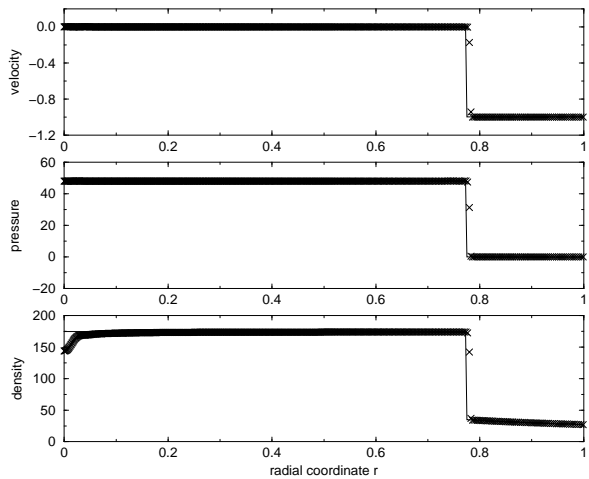,width=2.5in,angle=-90}}
\caption{Shock reflection test for an ultrarelativistic flow with
$u^{r} = -0.9999$c and $\rho_0 =8$ and EoS $p = 5 \times 10^{-4}
\rho^{\frac{5}{3}}$, which is reflected at the origin of the
coordinate system. We have plotted different fluid quantities at an
 evolution time $u=2.029$ as a function of the radial coordinate $r$. Top
panel: fluid velocity $u^r$. Middle Panel: pressure $p$. Bottom Panel:
rest mass density $\rho$. The solid line corresponds to the exact
solution, the crosses are taken from our numerical simulation.
For the above result, we made use of a non-equidistant radial grid 
$r = x/(1-x^{\frac{5}{2}})$ with 800 radial zones, the MC
slope limiter and the HLL approximate Riemann solver.}
\label{fig4.10}
\end{figure}

\subsection{Convergence tests}

We describe now some tests which check various properties
of spherically-symmetric core collapse. We choose a particular 
collapse model, for which the initial central density is 
$\rho_c=1.62 \times 10^{-8}$ (in units $G=c=M_{\odot}=1$), 
the polytropic constant is $\kappa = 0.46$, and the collapse 
is induced by resetting the adiabatic exponent to $\Gamma_1 = 
1.3$ (for the equilibrium model with $\Gamma=\frac{4}{3}$).
We use the hybrid EoS discussed in Section~\ref{subsec:EoS}.

\subsubsection{Thermal energy during the infall phase}

Before the central density of the collapsing core reaches nuclear 
densities, the collapse is exactly adiabatic. Hence, the thermal
energy, which vanishes initially, should vanish throughout this phase. 
This can be easily checked and used for convergence tests.
Fig.~\ref{fig4.11} shows the result after an integration time of
$30$ ms (when the central density has increased by roughly a factor 10). 
We find that the errors from the exact result $\epsilon_\mathrm{th} =0$
converge to zero, the convergence rate is 2. Note that although
$\epsilon_{th} \ge 0$ from the physical point of view, the numerical
errors can result in negative values for $\epsilon_\mathrm{th}$.

\begin{figure}[t]
\centerline{\psfig{file=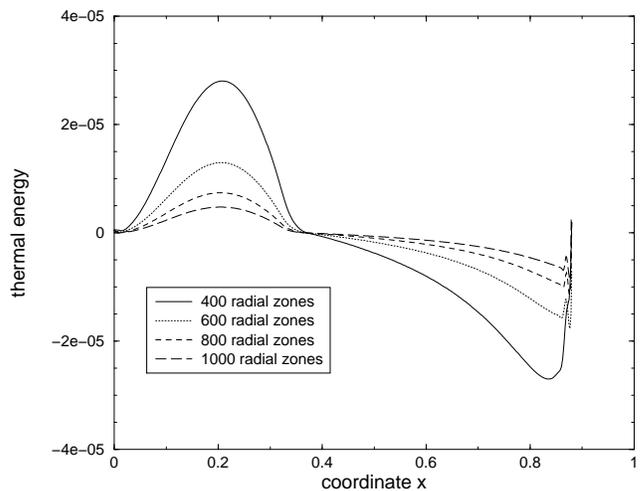,width=2.6in,angle=-90}}
\caption{Thermal energy as a function of the radial coordinate $x$
at $30$ ms for a compactified grid $r = \frac{200 x}{1-x^{2}}$ 
for different resolutions. Due to numerical errors, the thermal 
energy is different from zero. Deviations converge to zero, the 
convergence rate is 2.}
\label{fig4.11}
\end{figure}

\subsubsection{Time of bounce}

Using the axisymmetric code developed by Dimmelmeier et 
al.~\cite{DFM01,DFM02,DFM022} based on the
conformally flat metric approach, we can perform
comparisons between the evolutions of the same initial models.
As the conformally flat metric approximation is exact for spherical
models, comparisons in spherical symmetry are unambiguous.

\begin{figure}[t]
\centerline{\psfig{file=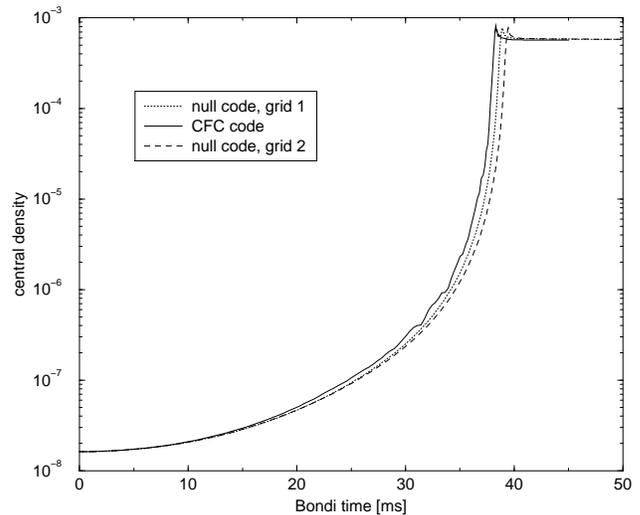,width=2.7in,angle=-90}}
\caption{Evolution of the central density for a core collapse
induced by resetting the adiabatic exponent to $\Gamma_1=1.30$.
The central density increases by almost 5 orders of magnitude, before 
the core bounces. Afterwards the central density stays almost constant. 
The different lines correspond to different grid functions and 
resolutions, see run 4, 1 and 5 in Table~\ref{tab:bounce}.}
\label{fig4.12}
\end{figure}

We define the time of bounce as the time, when the central density
reaches its maximum.  In order to start with the same initial data 
we initiate the collapse by ray-tracing the evolution of Dimmelmeier's 
code to obtain the initial data on our null cone. There is no principal 
advantage in starting with initial data on a null cone or on a Cauchy 
slice. Ideally, results from stellar evolution would give exact initial 
conditions for the core collapse, thus eliminating the artificial 
procedure of resetting $\Gamma$ to initiate the collapse. Fig.~\ref{fig4.12} 
shows the evolution of the central density for the relativistic code 
of~\cite{DFM02} and the results of our null code for two different 
grid functions. Table~\ref{tab:bounce} summarizes our results for the 
time of bounce.      

Assuming our code is exactly second order convergent and extrapolating
our results to an hypothetical infinite resolution, we
obtain from runs 3 and 4, that the infinite resolution run bounces
after $38.65$ ms.  This is internally consistent, a comparison of 
runs 2 and 4 results in a value of $38.66$ ms. Using an even higher resolution
for a different grid function in run 5, we observe a time of bounce close to
the converged result. Our results on the time of bounce are in very good 
agreement with the result of~\cite{DFM02}, who find a value of $38.32$ ms.
The observed difference of less than $1 \%$ is either due
to the fact that the result of~\cite{DFM02} is not converged, or due
to the different radial coordinates used in both codes, and thus small
differences in the initial data.

\begin{table}[htpb]
\centering
\caption{Times of bounce for different grid functions and
resolutions.}
\medskip
\begin{tabular}{c|c|c|c||c}
\hline
& code & grid     & radial     & time of \\
&      & function & resolution & bounce [ms] \\
\hline \hline
1 & CFC code~\cite{DFM02} & see ~\cite{DFM02} & $80^{*}$ & 38.32 \\ 
\hline
2 & null code        & $r = \frac{150 x}{1-x^{4}}$ & 600 & 40.86 \\ 
\hline
3 & null code        & $r = \frac{150 x}{1-x^{4}}$ & 800 & 39.90 \\
\hline
4 & null code        & $r = \frac{150 x}{1-x^{4}}$ & 1000 & 39.45 \\
\hline
5 & null code        & $r = 100 \tan(\frac{\pi}{2}x)$ & 1200 & 38.92
\\
\hline
\end{tabular}\\
\vspace{0.3cm}
\footnotesize{${}^*$ This number for the
radial resolution cannot be directly compared to the values of our
code, as we resolve the exterior vacuum region up to future null
infinity with our code as well.}
\label{tab:bounce}
\end{table}

As it can be seen in Fig.~\ref{fig4.12}, the comparison does not
only give very good agreement for the time of bounce, but also for
the dynamics of the central density in general. This is very
important, since it shows that the global dynamics of the core
collapse is correctly described in our numerical implementation.


\begin{thebibliography}{39}
\bibitem[{\citenamefont{note}(2003)}]{note1}
\bibinfo{note}{We use this term loosely, without claiming that we model the
  microphysics realistically.}
\bibitem[{\citenamefont{note}(2003)}]{note2}
\bibinfo{note}{Our notation follows the work~\cite{MSM91}. $E2$ denotes the
  electric part, $20$ denotes the $l=2$, $m=0$ quadrupolar part in an expansion
  of the gravitational wave strain in tensor harmonics.}
\bibitem[{\citenamefont{note}(2003)}]{note3}
\bibinfo{note}{For a resolution $(N_x,N_y) = (600,12)$, one time step is
  accomplished in about 2 s of CPU time on the DEC-Alpha workstations where we
  run the simulations achieving a performance of several hundred MFlops. Taking
  into account that about $7 \times 10^{5}$ time steps are needed to cover the
  evolution up to $u_B=50$ ms for this given resolution, one single simulation
  takes about 16 days.}
\end{thebibliography}
\end{document}